\documentclass[proceedings]{JHEP3}
\usepackage{amsfonts}
\usepackage{amsmath}
\usepackage{epsfig,multicol}

\newbox\mybox

\newcommand\fverb{\setbox\mybox=\hbox\bgroup\verb}
\newcommand\fverbdo{\egroup\medskip\noindent\fbox{\unhbox\mybox}\ }
\newcommand\fverbit{\egroup\item[\fbox{\unhbox\mybox}]}
\conference{Antilinear deformations of Coxeter groups}
\abstract{We construct complex root spaces remaining invariant
under antilinear involutions related to all Coxeter groups. We provide
two alternative constructions: One is based on deformations of factors of
the Coxeter element and the other based on the deformation of the longest element of the Coxeter group. Motivated by the fact that
non-Hermitian Hamiltonians admitting an antilinear symmetry may be used
to define consistent quantum mechanical systems with real discrete 
energy spectra, we subsequently employ our constructions to formulate deformations of Coxeter models remaining invariant under these extended Coxeter groups. We provide explicit and generic solutions for the 
Schr\"{o}dinger equation of these models for the eigenenergies and corresponding wavefunctions. A new feature of these novel models is that when compared with the undeformed case their solutions are usually no longer singular for an exchange of an amount of particles less than the dimension of the representation space of the roots. The simultaneous scattering of all particles in the model leads to anyonic exchange factors for processes which have no analogue in the undeformed case.}

\title{Antilinear deformations of Coxeter groups, an application to Calogero
models}
\author{Andreas Fring and Monique Smith \\
Centre for Mathematical Science, City University London,\\
Northampton Square, London EC1V 0HB, UK\\
E-mail: a.fring@city.ac.uk , abbc991@city.ac.uk}

\input{tcilatex}

\begin{document}

\section{Introduction}

It is by now a widely, although not yet universally, accepted fact that
non-Hermitian Hamiltonians admitting an antilinear symmetry may be used to
define consistent quantum mechanical systems with real energy spectra. This
property can be traced back to Wigner's observation \cite{EW} that operators
invariant under antilinear transformations possess real eigenvalues when
their eigenfunctions also respect this symmetry. Particular examples of such
an operator and symmetry are for instance a Hamiltonian and $\mathcal{PT}$%
-symmetry, i.e. a simultaneous parity transformation $\mathcal{P}$ and time
reversal $\mathcal{T}$, respectively. Many aspects of the latter possibility
are very much explored at present, see e.g. \cite{Benderrev,Rev3}, as part
of an activity initiated a bit more than a decade ago \cite{Bender:1998ke}.
Tracing back more than fifty years in the mathematical literature are the
closely related and often synonymously used concepts of quasi-Hermiticity 
\cite{Dieu,Will,Urubu} and pseudo-Hermiticity \cite%
{pseudo1,pseudo2,Mostafazadeh:2002hb}.

Here we will mainly explore the possibilities arising from antilinear
symmetries in general, which are usually realized directly on the dynamical
variables in case of quantum mechanical models or on the fields in case of
quantum field theories. There exist many models formulated generically in
terms of root systems in which the dynamical variables or fields lie in the
dual space of the roots, such as Calogero-Moser-Sutherland models, e.g. \cite%
{OP2} or Toda field theories, e.g. \cite{Wilson,DIO}, respectively. Since it
is usually quite difficult to identify the symmetry on the level of the
variables or fields in a controlled manner, the natural question arises
whether it is possible to have a more systematic construction and realize
the symmetry directly on the level of the roots, ideally in a completely
generic way that is irrespective of a specific root system and also
independently of a particular representation for the roots.

The first part of the manuscript, i.e. section 2, is devoted to the
development of the mathematical structure and framework where we will
provide generic deformations of root systems invariant under Coxeter
transformations, crystallographic as well as non-crystallographic ones. Our
starting point is to identify involutory maps inside the Coxeter groups and
deform them in such a way that they become antilinear. Possible candidates
are the Weyl reflections associated to each simple root, the factors of the
Coxeter element consisting of commuting Weyl reflections and the longest
element. For each of these options we set up a system of constraints for the
transformation matrix which maps simple roots into their complex
deformations. Subsequently we solve these sets of constraints on a
case--by-case basis for all Coxeter groups. Exploiting the fact that some
Coxeter groups are embedded into larger ones we also construct additional
complex solutions by means of the so-called folding procedure.

The second part of the manuscript is devoted to the application of our
constructions to the formulation and study of new models of Calogero type.
So far three non-equivalent methods have been explored to deform Calogero
models form real to complex systems: i) to add a $\mathcal{PT}$-invariant
term to the Hamiltonian \cite{Basu-Mallick:2000af,AF}, ii) to deform the
real Calogero model \cite{Milos,Jain,FZ} or iii) by constraining or
deforming real field equations, like the Boussinesq equation \cite%
{Assis:2009gt}, such that the poles of their solutions will give rise to
complex Calogero particles. With regard to i) it was shown in \cite{AF} that
for the terms added so far, the \textquotedblleft new\textquotedblright\
models simply correspond to the original ones with shifted momenta and
re-defined coupling constants. The second possibility ii) was first explored
in\ \cite{Milos}, where the deformation was carried out on the level of the
dynamical variables for an explicit representation of the $A_{2}$-root
system. In \cite{FZ} it was shown that these deformations could be
understood generically in the dual space, i.e. directly on the $A_{2}$-root
space. A construction for the $G_{2}$-root spaces and the corresponding
Calogero models was also provided in \cite{FZ}, whereas the $B_{2}$-case was
reported in \cite{Assis:2009gt}. It is this construction we generalize to
all Coxeter groups in this manuscript, albeit it turns out that the original
suggestion based on the deformation of the individual Weyl reflections is
too restrictive and may only be carried out for groups of rank 2. Instead
this construction needs to be viewed as a special case of a construction
based on the deformation factors of the Coxeter element consisting of
commuting Weyl reflections. For the antilinearly deformed Calogero
Hamiltonian we derive some solutions for the Schr\"{o}dinger equation for
the eigenenergies and wavefunctions. Our solutions are generalizations of
previously constructed ones in the sense that they are formulated in terms
of general roots, that is representation independent and irrespective of a
specific Coxeter group. Our derivation of the solutions is based on some
general identities which we present in appendix A, together with some
evidence of their validity. For self-consistency and easy reference we
present some case-by-case data for Coxeter groups in appendix B. Possibility
iii) is not yet formulated in a generic way in terms of root systems and we
will therefore not comment on it here.

\section{Root spaces invariant under antilinear involutions}

Before considering concrete physical models we will first provide the
general mathematical framework, which may also be applied to a different
physical setting than the one considered here. Our main aim in this section
is to construct complex extended root systems $\tilde{\Delta}(\varepsilon )$
which remain invariant under a newly defined antilinear involutary map. Our
starting point is to deform the real roots $\alpha _{i}\in \Delta \subset 
\mathbb{R}^{n}$ and seek to represent them in a complex space depending on
some deformation parameter $\varepsilon \in \mathbb{R}$ as $\tilde{\alpha}%
_{i}(\varepsilon )\in \tilde{\Delta}(\varepsilon )\subset $ $\mathbb{R}%
^{n}\oplus \imath \mathbb{R}^{n}$. For this purpose we define a linear
deformation map 
\begin{equation}
\delta :~\Delta \rightarrow \tilde{\Delta}(\varepsilon ),  \label{d}
\end{equation}%
relating simple roots $\alpha $ and deformed simple roots $\tilde{\alpha}$
as 
\begin{equation}
\alpha \mapsto \tilde{\alpha}=\theta _{\varepsilon }\alpha ,  \label{roots}
\end{equation}%
with the property that the new root space $\tilde{\Delta}$ remains invariant
under some antilinear involutory map $\omega $, i.e. $\omega :\tilde{\alpha}%
=\mu _{1}\alpha _{1}+\mu _{2}\alpha _{2}\mapsto \mu _{1}^{\ast }\omega
\alpha _{1}+\mu _{2}^{\ast }\omega \alpha _{2}$ for $\mu _{1}$, $\mu _{2}\in 
\mathbb{C}$, $\omega ^{2}=\mathbb{I}$ and $\omega :$ $\tilde{\Delta}%
\rightarrow \tilde{\Delta}$. To achieve this there are clearly various
possibilities conceivable. Here we make similar, albeit less constraining,
demands as in \cite{FZ} allowing for a generalization to root spaces
invariant under complex extended Coxeter groups $\mathcal{W}$ related to all
groups.

\subsection{$\mathcal{PT}$-symmetrically deformed Coxeter group factors\label%
{secPTCox}}

We wish to maintain the property that the entire deformed root space $\tilde{%
\Delta}(\varepsilon )$ can be generated analogously to the undeformed one $%
\Delta $, namely by some consecutive action of a deformed version of a
Coxeter element $\sigma \in \mathcal{W}$ on simple roots. The latter are
build up from a product of $\ell $ simple Weyl reflections 
\begin{equation}
\sigma _{i}(x):=x-2\frac{x\cdot \alpha _{i}}{\alpha _{i}^{2}}\alpha
_{i},\qquad \text{with \ \ \ }1\leq i\leq \ell \equiv \limfunc{rank}\mathcal{%
W},
\end{equation}%
that is 
\begin{equation}
\sigma =\prod\limits_{i=1}^{\ell }\sigma _{i}.
\end{equation}%
Since reflections do in general not commute, a Coxeter element is only
defined up to conjugation and therefore not unique. One way to fix ones
conventions is achieved by associating values $c_{i}=\pm 1$ to the vertices
of the Coxeter graphs, in such a way that no two vertices with the same
values are linked together. Consequently the simple roots associated to the
vertices split into two disjoints sets, say $V_{\pm }$, such that the
Coxeter element can be defined uniquely as 
\begin{equation}
\sigma :=\sigma _{-}\sigma _{+},\qquad \text{with }\sigma _{\pm
}:=\prod\limits_{i\in V_{\pm }}\sigma _{i},  \label{spm}
\end{equation}%
see e.g. \cite{Hum,Hum2,Moody,PD1,Mass2,Bradenno} for more details. Since
all elements in the same set commute, i.e. $[\sigma _{i},\sigma _{j}]=0$ for 
$i,j\in V_{+}$ or $i,j\in V_{-}$, the only ambiguity left at this stage is
the ordering between the $\sigma _{+}$ and $\sigma _{-}$.

In a general sense we explore here the possibility to identify the
antilinear map $\omega $ with a $\mathcal{PT}$-symmetry. In this concrete
setting the first attempts to pursue this idea to construct complex extended
root spaces were made in \cite{FZ}, where the authors used the fact that $%
\sigma _{i}^{2}=\mathbb{I}$ and identified the Weyl reflections as parity
transformations across all hyperplanes separating the Weyl chambers in the
rootspace. All these reflections were consistently deformable for $\mathcal{W%
}=A_{2},G_{2}$ \cite{FZ} and $\mathcal{W}=B_{2}$ \cite{Assis:2009gt}. For
groups with higher rank this amounts to a large number of constraints, which
are difficult to solve and might not even have a solution, as we shall see
indeed in section \ref{secPTWeyl}. Nonetheless, for the applications we have
in mind, i.e. to guarantee the reality of the spectra of some physical
operators as outlined above, one single deformed involutory map is in fact
sufficient.

Proceeding in this spirit with Weyl reflections leaves the question of which
one to choose as none is particularly distinct. However, there are some very
distinguished involutory elements contained in $\mathcal{W}$ of different
type, such as the aforementioned (\ref{spm}) two factors of the Coxeter
element $\sigma _{-}$ and $\sigma _{+}$, both satisfying $\sigma
_{-}^{2}=\sigma _{+}^{2}=\mathbb{I}$. We identify them here as parity
transformations and employ them to define two $\mathcal{PT}$-type operators
in two alternative ways 
\begin{equation}
\sigma _{\pm }^{\varepsilon }:=\theta _{\varepsilon }\sigma _{\pm }\theta
_{\varepsilon }^{-1}=\tau \sigma _{\pm }:\quad \tilde{\Delta}(\varepsilon
)\rightarrow \tilde{\Delta}(\varepsilon ),  \label{c1}
\end{equation}%
where $\tau $ mimics the time-reversal simply acting as a complex
conjugation. The operator $\theta _{\varepsilon }$ constitutes a realization
of the deformation map $\delta $ in (\ref{d}) relating deformed and
undeformed roots as specified in (\ref{roots}). The deformed Coxeter element
is then naturally defined as 
\begin{equation}
\sigma _{\varepsilon }:=\theta _{\varepsilon }\sigma \theta _{\varepsilon
}^{-1}=\sigma _{-}^{\varepsilon }\sigma _{+}^{\varepsilon }=\tau \sigma
_{-}\tau \sigma _{+}=\tau ^{2}\sigma _{-}\sigma _{+}=\sigma :\quad \tilde{%
\Delta}(\varepsilon )\rightarrow \tilde{\Delta}(\varepsilon ),  \label{ss}
\end{equation}%
i.e. it acts on $\tilde{\Delta}(\varepsilon )$ in the same way as $\sigma $
on $\Delta $. This means that the Coxeter transformation and the deformation
map commute 
\begin{equation}
\left[ \sigma ,\theta _{\varepsilon }\right] =0.  \label{c2}
\end{equation}%
Notice that from (\ref{c2}) follows that one equation in (\ref{c1}) implies
the other, i.e. the deformation of $\sigma _{+}$ yields the deformation of $%
\sigma _{-}$ and vice versa. The entire deformed root space $\tilde{\Delta}%
(\varepsilon )$ can be constructed in analogy to the undeformed case by
defining the quantity $\tilde{\gamma}_{i}=c_{i}\tilde{\alpha}_{i}$, which
serves as a representant for the deformed Coxeter orbit 
\begin{equation}
\Omega _{i}^{\varepsilon }:=\left\{ \tilde{\gamma}_{i},\sigma _{\varepsilon }%
\tilde{\gamma}_{i},\sigma _{\varepsilon }^{2}\tilde{\gamma}_{i},\ldots
,\sigma _{\varepsilon }^{h-1}\tilde{\gamma}_{i}\right\} =\theta
_{\varepsilon }\Omega _{i},  \label{omega}
\end{equation}%
such that 
\begin{equation}
\tilde{\Delta}(\varepsilon ):=\bigcup\limits_{i=1}^{\ell }\Omega
_{i}^{\varepsilon }=\theta _{\varepsilon }\Delta (\varepsilon )\text{.}
\label{omega2}
\end{equation}%
Note it is not enough to act just on the $\tilde{\alpha}_{i}$ to generate
the entire root space, but dressing them with the colour value will be
sufficient \cite{Hum,Hum2,Moody,PD1,Mass2,Bradenno}. Evidently when defining 
$\tilde{\Delta}(\varepsilon )$ in this way it will remain invariant under
the action of the (deformed) Coxeter element $\sigma _{\varepsilon }:\tilde{%
\Delta}(\varepsilon )\rightarrow \tilde{\Delta}(\varepsilon )$ and paramount
to our intentions the deformed root spaces are $\mathcal{PT}$-symmetric,
that is invariant with respect to the action of our map defined in (\ref{c1}%
) 
\begin{equation}
\sigma _{\pm }^{\varepsilon }:\tilde{\Delta}(\varepsilon )\rightarrow \theta
_{\varepsilon }\sigma _{\pm }\theta _{\varepsilon }^{-1}\tilde{\Delta}%
(\varepsilon )=\theta _{\varepsilon }\sigma _{\pm }\Delta (\varepsilon
)=\theta _{\varepsilon }\Delta (\varepsilon )=\tilde{\Delta}(\varepsilon ).
\label{a}
\end{equation}%
This observation suggests to demand a one-to-one relation between the
individual roots, such that $\tilde{\Delta}(\varepsilon )$ is isomorphic to $%
\Delta $. This is guaranteed with the limit 
\begin{equation}
\lim_{\varepsilon \rightarrow 0}\tilde{\alpha}_{i}(\varepsilon )=\alpha _{i},
\label{c4}
\end{equation}%
and therefore we have the reduction $\lim_{\varepsilon \rightarrow 0}\tilde{%
\Delta}(\varepsilon )=\Delta $, i.e. $\lim_{\varepsilon \rightarrow 0}\theta
_{\varepsilon }=\mathbb{I}$.

In principle, provided $\theta _{\varepsilon }$ can be constructed, this
will allow us already to formulate new $\mathcal{PT}$-symmetric physical
models based on roots by means of the deformation map $\delta :\alpha
\mapsto \tilde{\alpha}(\varepsilon )$. However, the number of free
parameters is still very large and it is natural to impose further
constraints. Motivated by the physical applications we have in mind, we
would like the kinetic energy term and possibly other terms in the models to
remain invariant under the deformation. This will be guaranteed when we
demand the invariance of the inner products in the corresponding root spaces 
\begin{equation}
\alpha _{i}\cdot \alpha _{j}=\tilde{\alpha}_{i}\cdot \tilde{\alpha}_{j}.
\label{scalar}
\end{equation}
This means $\theta _{\varepsilon }$ is an isometry and we demand therefore 
\begin{equation}
\theta _{\varepsilon }^{\ast }=\theta _{\varepsilon }^{-1}\text{\qquad
and\qquad }\det \theta _{\varepsilon }=\pm 1.  \label{c3}
\end{equation}

In summary, it turns out that given $\sigma _{+}$ or $\sigma _{-}$ for a
particular Coxeter group $\mathcal{W}$, we can construct their $\mathcal{PT}$%
-symmetric, or better antilinear, deformations by solving the constraints (%
\ref{c1}), (\ref{c2}), (\ref{c3}) and (\ref{c4}) that is 
\begin{equation}
\theta _{\varepsilon }^{\ast }\sigma _{\pm }=\sigma _{\pm }\theta
_{\varepsilon },\quad \left[ \sigma ,\theta _{\varepsilon }\right] =0,\quad
\theta _{\varepsilon }^{\ast }=\theta _{\varepsilon }^{-1},\quad \det \theta
_{\varepsilon }=\pm 1\quad \text{and\quad }\lim_{\varepsilon \rightarrow
0}\theta _{\varepsilon }=\mathbb{I}\text{,}  \label{const}
\end{equation}%
for $\theta _{\varepsilon }$. Up to a certain point we demonstrate this now
for all Coxeter groups.

In the light of the fact that the Coxeter element commutes with $\theta
_{\varepsilon }$ and the last relation in (\ref{const}), we make the Ansatz 
\begin{equation}
\theta _{\varepsilon }=\sum\limits_{k=0}^{h-1}c_{k}(\varepsilon )\sigma
^{k},\qquad \text{with }\lim_{\varepsilon \rightarrow 0}c_{k}(\varepsilon
)=\left\{ 
\begin{array}{l}
1\quad k=0 \\ 
0\quad k\neq 0%
\end{array}%
\right. ,~c_{k}(\varepsilon )\in \mathbb{C}.  \label{Ansatz}
\end{equation}%
Next we try to satisfy the first relation in (\ref{const}). Using the
relations $\sigma _{-}\sigma ^{-1}=\sigma \sigma _{-}$ and $\sigma ^{h}=1$,
we obtain with (\ref{Ansatz}) 
\begin{equation}
\theta _{\varepsilon }^{\ast }\sigma _{-}=\sum\limits_{k=0}^{h-1}c_{k}^{\ast
}(\varepsilon )\sigma ^{k}\sigma _{-}=\sum\limits_{k=0}^{h-1}c_{k}^{\ast
}(\varepsilon )\sigma _{-}\sigma ^{h-k},
\end{equation}%
which equals 
\begin{equation}
\sigma _{-}\theta _{\varepsilon }=\sum\limits_{k=0}^{h-1}c_{k}(\varepsilon
)\sigma _{-}\sigma ^{k},
\end{equation}%
when 
\begin{equation}
c_{h-k}(\varepsilon )=c_{k}^{\ast }(\varepsilon ).  \label{C}
\end{equation}%
As we expect from the comment after (\ref{c2}), the constraint $\theta
_{\varepsilon }^{\ast }\sigma _{+}=\sigma _{+}\theta _{\varepsilon }$ yields
the same relation (\ref{C}) where in the derivation we have to use, however, 
$\sigma _{+}\sigma =\sigma ^{-1}\sigma _{+}$ instead. Since $%
c_{h}(\varepsilon )=c_{0}(\varepsilon )$ the equality (\ref{C}) implies that 
$c_{0}(\varepsilon )=:r_{0}(\varepsilon )\in \mathbb{R}$ and furthermore we
deduce that $c_{h/2}(\varepsilon )=:r_{h/2}(\varepsilon )\in \mathbb{R}$
when $h$ is even. Furthermore, we may take the $c_{k}(\varepsilon )$ to be
of the form $c_{k}(\varepsilon )=\imath r_{k}(\varepsilon )$. Mostly it
turns out that $r_{k}(\varepsilon )$ $\in \mathbb{R}$, but we will not
assume this from the start as we do not wish to exclude possible solutions.
Therefore we can write 
\begin{equation}
\theta _{\varepsilon }=\left\{ 
\begin{array}{ll}
r_{0}(\varepsilon )\mathbb{I}+\imath
\sum\limits_{k=1}^{(h-1)/2}r_{k}(\varepsilon )(\sigma ^{k}-\sigma ^{-k}) & 
\text{for }h\text{ odd,} \\ 
r_{0}(\varepsilon )\mathbb{I}+r_{h/2}(\varepsilon )\sigma ^{h/2}+\imath
\sum\limits_{k=1}^{h/2-1}r_{k}(\varepsilon )(\sigma ^{k}-\sigma ^{-k})~~~ & 
\text{for }h\text{ even.}%
\end{array}%
\right.
\end{equation}%
We can easily diagonalize $\theta _{\varepsilon }$ by recalling \cite{Hum2}
the eigenvalue equation for the Coxeter element 
\begin{equation}
\sigma v_{n}=e^{2\pi \imath s_{n}/h}v_{n},  \label{exp}
\end{equation}%
with $s_{n}$ being the exponents of a particular Coxeter group $\mathcal{W}$%
, see appendix B for explicit values. Defining the matrix $\vartheta
=\{v_{1},v_{2},\ldots ,v_{\ell }\}$, we diagonalize the Coxeter element
simply as $\sigma =\vartheta \hat{\sigma}\vartheta ^{-1}$ with $\hat{\sigma}%
_{nn}=e^{2\pi \imath s_{n}/h}$, such that the deformation matrix
diagonalizes as 
\begin{equation}
\theta _{\varepsilon }=\vartheta \hat{\theta}_{\varepsilon }\vartheta ^{-1},
\label{tt}
\end{equation}%
with eigenvalues 
\begin{equation}
(\hat{\theta}_{\varepsilon })_{nn}=\left\{ 
\begin{array}{ll}
r_{0}(\varepsilon )-2\sum\limits_{k=1}^{(h-1)/2}r_{k}(\varepsilon )\sin
\left( \frac{2\pi k}{h}s_{n}\right) & \text{for }h\text{ odd,} \\ 
r_{0}(\varepsilon )+(-1)^{s_{n}}r_{h/2}(\varepsilon
)-2\sum\limits_{k=1}^{h/2-1}r_{k}(\varepsilon )\sin \left( \frac{2\pi k}{h}%
s_{n}\right) ~~~ & \text{for }h\text{ even.}%
\end{array}%
\right.
\end{equation}%
This means that the constraint $\det \theta _{\varepsilon }=\pm 1$ in (\ref%
{const}) is equivalent to $\det \hat{\theta}_{\varepsilon }=\pm 1$ and
therefore 
\begin{equation}
\begin{array}{ll}
\pm 1=\prod\limits_{n=1}^{\ell }\left[ r_{0}(\varepsilon
)-2\sum\limits_{k=1}^{(h-1)/2}r_{k}(\varepsilon )\sin \left( \frac{2\pi k}{h}%
s_{n}\right) \right] & \text{for }h\text{ odd,} \\ 
\pm 1=\prod\limits_{n=1}^{\ell }\left[ r_{0}(\varepsilon
)+(-1)^{n}r_{h/2}(\varepsilon )-2\sum\limits_{k=1}^{h/2-1}r_{k}(\varepsilon
)\sin \left( \frac{2\pi k}{h}s_{n}\right) \right] ~~~ & \text{for }h\text{
even.}%
\end{array}
\label{e1}
\end{equation}%
Next we implement the third relation in (\ref{const}), which, using (\ref{tt}%
), corresponds to the $\ell $ equations 
\begin{equation}
\vartheta ^{-1}\vartheta ^{\ast }\hat{\theta}_{\varepsilon }(\vartheta
^{\ast })^{-1}\vartheta =\hat{\theta}_{\varepsilon }^{-1}.  \label{e2}
\end{equation}%
What is left is to find are the $(h-1)/2$ or $h/2+1$ unknown functions $%
r_{i}(\varepsilon )$ when $h$ is odd or even, respectively, from the $\ell
+1 $ equations (\ref{e1}) and (\ref{e2}). We carry out this task
case-by-case for specific Coxter groups in section \ref{casebycase}.

\subsection{$\mathcal{CT}$-symmetrically deformed longest element \label%
{secCT}}

Intuitively it would be more natural to have just one deformed involutory
map from the start instead of two. In fact there exist one very distinct
involution in $\mathcal{W}$, called the longest element. The length of an
element in the Coxeter group $\mathcal{W}$ is defined as the smallest amount
of simple Weyl reflections $\sigma _{i}$ needed to express that element, see
e.g. \cite{Hum2}. Since Coxeter groups are finite, there exists an element
in $\mathcal{W}$ of maximal length, i.e. the longest element, which we
denote as $w_{0}$. The length of this element equals the number of positive
roots $h\ell $, with $h$ being the Coxeter number of $\mathcal{W}$. The map $%
w_{0}$ is involutive, mapping the set of positive roots $\Delta _{+}\subset $
$\mathbb{R}^{n}$ to negative ones $\Delta _{-}$ $\subset $ $\mathbb{R}^{n}$
and vice versa 
\begin{equation}
w_{0}:~~\Delta _{\pm }\rightarrow \Delta _{\mp },  \label{1}
\end{equation}%
where $w_{0}^{2}=\mathbb{I}$. Two specific simple roots, say $\alpha _{i}$
and $\alpha _{\bar{\imath}}$, are linearly related by $w_{0}$ as 
\begin{equation}
\alpha _{i}\mapsto -\alpha _{\bar{\imath}}=(w_{0}\alpha )_{i}.  \label{map}
\end{equation}%
Here we have borrowed the notation from the context of affine Toda field
theories, where it was found \cite{Bradenno} that the longest element serves
as charge conjugation operator $\mathcal{C}$, mapping a particle of type $i$
to its anti-particle $\bar{\imath}$. From a more mathematical perspective
this map is a particular symmetry of the Dynkin diagrams, see e.g. \cite{DIO}%
.

\unitlength=0.6500000pt 
\begin{picture}(300.0,70.00)(150.00,125.00)

\put(140.00,150.00){\makebox(0.00,0.00){$A_{\ell}:$}}
\put(389.00,165.00){\makebox(0.00,0.00){ ${\alpha}_{\ell}$ }}
\put(348.0,165.00){\makebox(0.00,0.00){${\alpha}_{\ell-1}$}}
\put(296.00,165.00){\makebox(0.00,0.00){${\alpha}_3$}}
\put(255.00,165.00){\makebox(0.00,0.00){${\alpha}_2$}}
\put(215.00,165.00){\makebox(0.00,0.00){${\alpha}_1$}}
\put(321.00,150.00){\makebox(0.00,0.00){$\cdots$}}

\put(350.00,150.00){\line(1,0){30.00}}
\put(336.00,150.00){\line(1,0){5.00}}
\put(300.00,150.00){\line(1,0){5.00}}
\put(260.00,150.00){\line(1,0){30.00}}
\put(220.00,150.00){\line(1,0){30.00}}

\put(215.00,150.00){\circle*{10.00}}
\put(255.00,150.00){\circle*{10.00}}
\put(295.00,150.00){\circle*{10.00}}
\put(345.00,150.00){\circle*{10.00}}
\put(385.00,150.00){\circle*{10.00}}
\put(432.00,157.00){$-w_0$}
\put(440.00,147.00){$ \longrightarrow $}

\put(675.00,165.00){\makebox(0.00,0.00){ ${\alpha}_{1}$ }}
\put(635.00,165.00){\makebox(0.00,0.00){ ${\alpha}_{2}$ }}
\put(595.00,165.00){\makebox(0.00,0.00){ ${\alpha}_{3}$ }}
\put(545.00,165.00){\makebox(0.00,0.00){ ${\alpha}_{\ell-1}$ }}
\put(505.00,165.00){\makebox(0.00,0.00){ ${\alpha}_{\ell}$ }}
\put(571.00,150.00){\makebox(0.00,0.00){$\cdots$}}

\put(640.00,150.00){\line(1,0){30.00}}
\put(600.00,150.00){\line(1,0){30.00}}
\put(585.00,150.00){\line(1,0){5.00}}
\put(550.00,150.00){\line(1,0){5.00}}
\put(510.00,150.00){\line(1,0){30.00}}

\put(675.00,150.00){\circle*{10.00}}
\put(635.00,150.00){\circle*{10.00}}
\put(595.00,150.00){\circle*{10.00}}
\put(505.00,150.00){\circle*{10.00}}
\put(545.00,150.00){\circle*{10.00}}
\end{picture}

\unitlength=0.6500000pt 
\begin{picture}(300.0,70.00)(150.00,125.00)

\put(140.00,150.00){\makebox(0.00,0.00){$D_{2\ell + 1}:$}}
\put(401.00,180.00){\makebox(0.00,0.00){ ${\alpha}_{2\ell+1}$ }}
\put(396.00,120.00){\makebox(0.00,0.00){ ${\alpha}_{2\ell}$ }}
\put(373.0,150.00){\makebox(0.00,0.00){${\alpha}_{ 2\ell-1 }$}}
\put(296.00,165.00){\makebox(0.00,0.00){${\alpha}_3$}}
\put(255.00,165.00){\makebox(0.00,0.00){${\alpha}_2$}}
\put(215.00,165.00){\makebox(0.00,0.00){${\alpha}_1$}}
\put(321.00,150.00){\makebox(0.00,0.00){$\cdots$}}

\put(336.00,150.00){\line(1,0){5.00}}
\put(300.00,150.00){\line(1,0){5.00}}
\put(260.00,150.00){\line(1,0){30.00}}
\put(220.00,150.00){\line(1,0){30.00}}
\put(349.00,146.33){\line(1,-1){24.0}}
\put(371.67,176.67){\line(-1,-1){24.00}}

\put(215.00,150.00){\circle*{10.00}}
\put(255.00,150.00){\circle*{10.00}}
\put(295.00,150.00){\circle*{10.00}}
\put(345.00,150.00){\circle*{10.00}}
\put(375.00,180.00){\circle*{10.00}}
\put(375.00,120.00){\circle*{10.00}}
\put(432.00,157.00){$-w_0$}
\put(440.00,147.00){$ \longrightarrow $}
\put(687.00,180.00){\makebox(0.00,0.00){ ${\alpha}_{2\ell}$ }}
\put(691.00,120.00){\makebox(0.00,0.00){ ${\alpha}_{2\ell+1}$ }}
\put(663.0,150.00){\makebox(0.00,0.00){${\alpha}_{ 2\ell-1 }$}}
\put(586.00,165.00){\makebox(0.00,0.00){${\alpha}_3$}}
\put(545.00,165.00){\makebox(0.00,0.00){${\alpha}_2$}}
\put(505.00,165.00){\makebox(0.00,0.00){${\alpha}_1$}}
\put(611.00,150.00){\makebox(0.00,0.00){$\cdots$}}

\put(626.00,150.00){\line(1,0){5.00}}
\put(590.00,150.00){\line(1,0){5.00}}
\put(550.00,150.00){\line(1,0){30.00}}
\put(510.00,150.00){\line(1,0){30.00}}
\put(639.00,146.33){\line(1,-1){24.0}}
\put(661.67,176.67){\line(-1,-1){24.00}}

\put(505.00,150.00){\circle*{10.00}}
\put(545.00,150.00){\circle*{10.00}}
\put(585.00,150.00){\circle*{10.00}}
\put(635.00,150.00){\circle*{10.00}}
\put(665.00,180.00){\circle*{10.00}}
\put(665.00,120.00){\circle*{10.00}}

\end{picture}

\unitlength=0.6500000pt 
\begin{picture}(370.00,107.58)(180.00,0.00)
\put(170.00,38.00){\makebox(0.00,0.00){$E_6:$}}

\put(405.00,52.00){\makebox(0.00,0.00){$\alpha_6$}}
\put(365.00,52.00){\makebox(0.00,0.00){$\alpha_5$}}
\put(335.00,52.00){\makebox(0.00,0.00){$\alpha_4$}}
\put(285.00,52.00){\makebox(0.00,0.00){${\alpha}_3$}}
\put(325.00,90.00){\makebox(0.00,0.00){${\alpha}_2$}}
\put(245.00,52.00){\makebox(0.00,0.00){${\alpha}_1$}}

\put(370.00,38.00){\line(1,0){30.00}}
\put(330.00,38.00){\line(1,0){30.00}}
\put(325.33,72.67){\line(0,-1){31.00}}
\put(290.00,38.00){\line(1,0){30.00}}
\put(250.00,38.00){\line(1,0){30.00}}

\put(325.00,78.00){\circle*{10.00}}
\put(245.00,38.00){\circle*{10.00}}
\put(325.00,38.00){\circle*{10.00}}
\put(365.00,38.00){\circle*{10.00}}
\put(405.00,38.00){\circle*{10.00}}
\put(285.00,38.00){\circle*{10.00}}
\put(470.00,45.00){$-w_0$}
\put(478.00,35.00){$ \longrightarrow $}
\put(615.00,78.00){\circle*{10.00}}
\put(535.00,38.00){\circle*{10.00}}
\put(615.00,38.00){\circle*{10.00}}
\put(655.00,38.00){\circle*{10.00}}
\put(695.00,38.00){\circle*{10.00}}
\put(575.00,38.00){\circle*{10.00}}

\put(695.00,52.00){\makebox(0.00,0.00){$\alpha_1$}}
\put(655.00,52.00){\makebox(0.00,0.00){$\alpha_3$}}
\put(625.00,52.00){\makebox(0.00,0.00){$\alpha_4$}}
\put(575.00,52.00){\makebox(0.00,0.00){${\alpha}_5$}}
\put(615.00,90.00){\makebox(0.00,0.00){${\alpha}_2$}}
\put(535.00,52.00){\makebox(0.00,0.00){${\alpha}_6$}}

\put(660.00,38.00){\line(1,0){30.00}}
\put(620.00,38.00){\line(1,0){30.00}}
\put(615.33,72.67){\line(0,-1){31.00}}
\put(580.00,38.00){\line(1,0){30.00}}
\put(540.00,38.00){\line(1,0){30.00}}

\end{picture}

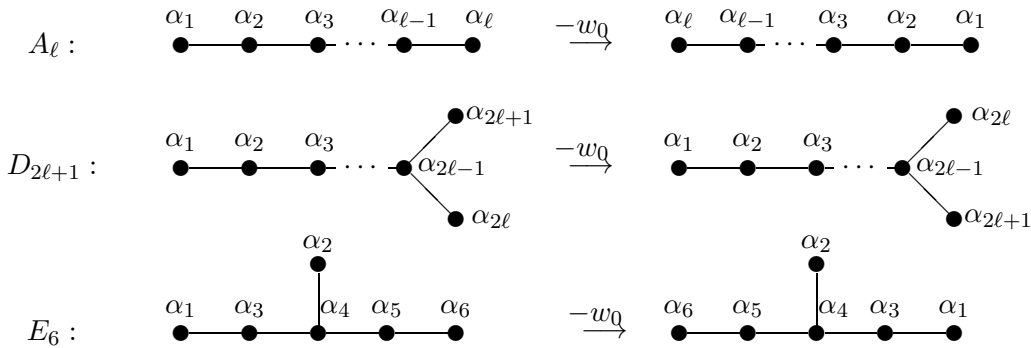
\smallskip \noindent {\small Figure 1: The action of $-w_{0}$ on the Dynkin
diagrams.}

The longest element admits a concrete realization in terms of products of
Coxeter transformations $\sigma $. The unique longest element can be
expressed as \cite{Bradenno} 
\begin{equation}
w_{0}=\left\{ 
\begin{array}{ll}
\sigma ^{h/2} & \quad \text{for }h\text{ even,} \\ 
\sigma _{+}\sigma ^{(h-1)/2} & \quad \text{for }h\text{ odd.}%
\end{array}%
\right.  \label{w0}
\end{equation}%
For the individual algebras the roots\ $\alpha _{\bar{\imath}}$ in (\ref{map}%
) are calculated directly or identified from the symmetries of the Dynkin
diagrams \cite{DIO} as 
\begin{equation}
\begin{array}{rl}
A_{\ell }: & \alpha _{\bar{\imath}}=\alpha _{\ell +1-i}, \\ 
D_{\ell }: & \left\{ 
\begin{array}{ll}
\alpha _{\bar{\imath}}=\alpha _{i}\text{ ~~for }1\leq i\leq \ell \text{, } & 
\text{when }\ell \text{ even} \\ 
\alpha _{\bar{\imath}}=\alpha _{i}\text{ \ \ for }1\leq i\leq \ell -2\text{, 
}\alpha _{\bar{\ell}}=\alpha _{\ell -1}\text{, } & \text{when }\ell \text{
odd,}%
\end{array}%
\right. \\ 
E_{6}: & \alpha _{\bar{1}}=\alpha _{6},\alpha _{\bar{3}}=\alpha _{5},\alpha
_{\bar{2}}=\alpha _{2},\alpha _{\bar{4}}=\alpha _{4}, \\ 
B_{\ell },C_{\ell },E_{7},E_{8},F_{4},G_{2}: & \alpha _{\bar{\imath}}=\alpha
_{i}.%
\end{array}
\label{case}
\end{equation}

Defining then a $\mathcal{CT}$-operator in analogy to (\ref{c1}) in two
alternative ways, we have 
\begin{equation}
w_{0}^{\varepsilon }=\theta _{\varepsilon }w_{0}\theta _{\varepsilon
}^{-1}=\tau w_{0}.  \label{w}
\end{equation}
When $\left[ \sigma ,\theta _{\varepsilon }\right] =0$ this equation has no
solution for even $h$, since $w_{0}^{\varepsilon }=\theta _{\varepsilon
}\sigma ^{h/2}\theta _{\varepsilon }^{-1}=\sigma ^{h/2}=\tau \sigma ^{h/2}$,
which is evidently a contradiction. Whereas for odd $h$ the realization (\ref%
{w0}) in (\ref{w}) yields $\theta _{\varepsilon }\sigma _{+}\sigma
^{(h-1)/2}\theta _{\varepsilon }^{-1}=\theta _{\varepsilon }\sigma
_{+}\theta _{\varepsilon }^{-1}\sigma ^{(h-1)/2}=\tau \sigma _{+}\sigma
^{(h-1)/2}$, which equals (\ref{c1}) when canceling $\sigma ^{(h-1)/2}$,
such that this case is equivalent to the one described in the previous
subsection. This means in order to obtain a new solution from (\ref{w}) we
need to assume $\left[ \sigma ,\theta _{\varepsilon }\right] \neq 0$.

This fact implies immediately that we have now two options to construct the
remaining nonsimple roots. We may either define in complete analogy to (\ref%
{omega}) and (\ref{omega2}) a root space which remains invariant under the
action of the deformed Coxeter transformation. This root space is then also $%
\mathcal{CT}$-symmetric 
\begin{equation}
w_{0}^{\varepsilon }:\tilde{\Delta}(\varepsilon )\rightarrow \theta
_{\varepsilon }w_{0}\theta _{\varepsilon }^{-1}\tilde{\Delta}(\varepsilon
)=\theta _{\varepsilon }w_{0}\Delta (\varepsilon )=\theta _{\varepsilon
}\Delta (\varepsilon )=\tilde{\Delta}(\varepsilon ).
\end{equation}
Alternatively we could also define 
\begin{equation}
\hat{\Omega}_{i}^{\varepsilon }:=\left\{ \tilde{\gamma}_{i},\sigma \tilde{%
\gamma}_{i},\sigma ^{2}\tilde{\gamma}_{i},\ldots ,\sigma ^{h-1}\tilde{\gamma}%
_{i}\right\}
\end{equation}
and the entire root space as $\tilde{\Delta}(\varepsilon
):=\bigcup\nolimits_{i=1}^{\ell }\hat{\Omega}_{i}^{\varepsilon }$. However,
this root space will only remain invariant under the action of $\sigma $
instead of $\sigma ^{\varepsilon }$ and in addition it will not be $\mathcal{%
CT}$-symmetric. This definition is therefore unsuitable for our purposes
here.

Using the two definitions in (\ref{w}) leads on one hand to 
\begin{equation}
w_{0}^{\varepsilon }\tilde{\alpha}=\theta _{\varepsilon }w_{0}\theta
_{\varepsilon }^{-1}\theta _{\varepsilon }\alpha =\theta _{\varepsilon
}w_{0}\alpha =-\theta _{\varepsilon }\bar{\alpha},
\end{equation}%
and on the other to 
\begin{equation}
w_{0}^{\varepsilon }\tilde{\alpha}=\tau w_{0}\tilde{\alpha}=-\tau \bar{%
\tilde{\alpha}}=-\bar{\tilde{\alpha}}^{\ast },  \label{es}
\end{equation}%
such that 
\begin{equation}
\left( \theta _{\varepsilon }\right) _{ij}=\left( \theta _{\varepsilon
}^{\ast }\right) _{\bar{\imath}\bar{\jmath}}.  \label{th}
\end{equation}%
As in the previous subsection we require the inner products to be preserved (%
\ref{const}), such that in summary the set of determining equations result
to 
\begin{equation}
\theta _{\varepsilon }^{\ast }w_{0}=w_{0}\theta _{\varepsilon },\quad \left[
\sigma ,\theta _{\varepsilon }\right] \neq 0,\quad \theta _{\varepsilon
}^{\ast }=\theta _{\varepsilon }^{-1},\quad \det \theta _{\varepsilon }=\pm
1\quad \text{and\quad }\lim_{\varepsilon \rightarrow 0}\theta _{\varepsilon
}=\mathbb{I}\text{.}  \label{constw}
\end{equation}

In this case it is instructive to separate $\theta _{\varepsilon }$ into its
real and imaginary part $\left( \theta _{\varepsilon }\right)
_{ij}=R_{i}^{j}(\varepsilon )+\imath I_{i}^{j}(\varepsilon )$ and therefore
expand an arbitrary simple deformed root in terms of the $\ell $ simple
roots as 
\begin{equation}
\tilde{\alpha}_{i}(\varepsilon ):=\sum\limits_{j=1}^{\ell }\left(
R_{i}^{j}(\varepsilon )\alpha _{j}+\imath I_{i}^{j}(\varepsilon )\alpha
_{j}\right) ,  \label{root}
\end{equation}%
with $R_{i}^{j}(\varepsilon )$ and $I_{i}^{j}(\varepsilon )$ being some real
valued functions satisfying 
\begin{equation}
\lim_{\varepsilon \rightarrow 0}R_{i}^{j}(\varepsilon )=\left\{ 
\begin{array}{c}
1\text{\qquad for }i=j \\ 
0\text{\qquad for }i\neq j%
\end{array}%
\right. \qquad \text{and\qquad }\lim_{\varepsilon \rightarrow
0}I_{i}^{j}(\varepsilon )=0.  \label{lim}
\end{equation}%
The relation (\ref{th}) then implies that 
\begin{equation}
R_{i}^{j}(\varepsilon )=R_{\bar{\imath}}^{\bar{\jmath}}(\varepsilon )\qquad 
\text{and\qquad }I_{i}^{j}(\varepsilon )=-I_{\bar{\imath}}^{\bar{\jmath}%
}(\varepsilon ).  \label{RI}
\end{equation}%
This means for Coxeter groups in which for all simple roots are
self-conjugate $\alpha _{i}=\alpha _{\bar{\imath}}$ a nontrivial complex $%
\mathcal{CT}$-symmetric deformation of the longest element can not exist.

\subsection{$\mathcal{PT}$-symmetrically deformed Weyl reflections \label%
{secPTWeyl}}

Yet another possibility would be to identify the parity operator $\mathcal{P}
$ with the Weyl reflections $\sigma _{i}$ across all hyperplanes separating
the Weyl chambers as suggested in \cite{FZ}. For rank $2$ Coxeter groups
this construction is identical to the one in section \ref{secPTCox} and the
natural question is whether it can be generalized to higher rank. We present
here a simple argument which proves that this is in fact not possible.

Assuming that we can consistently deform at least three Weyl reflections
according to 
\begin{equation}
\sigma _{i}^{\varepsilon }=\theta _{\varepsilon }\sigma _{i}\theta
_{\varepsilon }^{-1}=\tau \sigma _{i}\text{,}\quad \qquad \text{for }i=j,k,l,
\label{w1}
\end{equation}
it follows that 
\begin{equation}
\sigma _{j}^{\varepsilon }\sigma _{k}^{\varepsilon }\sigma _{l}^{\varepsilon
}=\theta _{\varepsilon }\sigma _{j}\sigma _{k}\sigma _{l}\theta
_{\varepsilon }^{-1}=\tau ^{3}\sigma _{j}\sigma _{k}\sigma _{l}=\tau \sigma
_{j}\sigma _{k}\sigma _{l}.  \label{w2}
\end{equation}
Demanding that inner products are preserved, we may employ (\ref{c3}) and
combine it with (\ref{w1}) to derive $\sigma _{i}=\theta _{\varepsilon
}\sigma _{i}\theta _{\varepsilon }$. Therefore we have 
\begin{equation}
\sigma _{j}\sigma _{k}\sigma _{l}=\theta _{\varepsilon }\sigma _{j}\theta
_{\varepsilon }\sigma _{k}\theta _{\varepsilon }\sigma _{l}\theta
_{\varepsilon }=\theta _{\varepsilon }\sigma _{j}\sigma _{k}\sigma
_{l}\theta _{\varepsilon },  \label{w3}
\end{equation}
which together with (\ref{w2}) implies that 
\begin{equation}
\tau =\theta _{\varepsilon }^{2}.  \label{tq}
\end{equation}
As this is impossible to solve this means more than two Weyl reflections can
not be consistently $\mathcal{PT}$-deformed in a simultaneous manner.

\subsection{Deformed root systems, case-by-case solutions\label{casebycase}}

On a case-by case basis for individual Coxeter groups we will now provide
explicit solutions for the set of constraining equations for deformed root
systems based on antilinear deformations of the $\sigma _{\pm }$ as outlined
in section \ref{secPTCox} and where possible also based on deformations of
the longest element $w_{0}$ as explained in section \ref{secCT}.

\subsubsection{$\tilde{\Delta}(\protect\varepsilon )$ for $A_{\ell}$}

Our convention for the labelling of the roots is depicted in figure 1.

\noindent\textbf{{$\tilde{\Delta}(\varepsilon )$ for $A_{2}$}}

\noindent As explained after equation (\ref{w}), we should obtain identical
deformed root spaces from the two different construction methods in this
case as the Coxeter number $h$ is odd .

\subparagraph{\noindent $\mathcal{CT}$-symmetrically deformed longest element%
}

Let us start with the construction of a $\mathcal{CT}$-symmetric deformation
of $w_{0}$. According to (\ref{case}) in the $A_{2}$-case the two simple
roots are related to each other by the longest element or in affine Toda
particle terminology they are conjugate to each other, i.e. $\bar{1}=2$.
Using the expansion (\ref{root}) and the constraints (\ref{RI}) the deformed
roots acquire the form 
\begin{eqnarray}
\tilde{\alpha}_{1} &=&R_{1}^{1}(\varepsilon )\alpha
_{1}+R_{1}^{2}(\varepsilon )\alpha _{2}+\imath (I_{1}^{1}(\varepsilon
)\alpha _{1}+I_{1}^{2}(\varepsilon )\alpha _{2}),  \label{a1} \\
\tilde{\alpha}_{2} &=&R_{1}^{2}(\varepsilon )\alpha
_{1}+R_{1}^{1}(\varepsilon )\alpha _{2}-\imath (I_{1}^{2}(\varepsilon
)\alpha _{1}+I_{1}^{1}(\varepsilon )\alpha _{2}).  \label{a2}
\end{eqnarray}%
Demanding next that the inner products are preserved (\ref{scalar}) amounts
to three further constraint, such that the four free functions in (\ref{a1}%
), (\ref{a2}) are reduced to only one. We obtain the two solutions 
\begin{equation}
R_{1}^{2}=0,\quad I_{1}^{2}=2I_{1}^{1},\quad \left( R_{1}^{1}\right) ^{2}-%
\frac{3}{4}\left( I_{1}^{2}\right) ^{2}=1\quad \text{and\quad }%
1\leftrightarrow 2.  \label{cc}
\end{equation}%
The third relation in (\ref{cc}) is solved for instance by $R_{1}^{1}=\cosh
\varepsilon $, $I_{1}^{2}=2/\sqrt{3}\sinh \varepsilon $ satisfying also the
limiting constraint (\ref{lim}) for $\varepsilon \rightarrow 0$.
Accordingly, the deformed simple roots are 
\begin{eqnarray}
\tilde{\alpha}_{1} &=&\cosh \varepsilon \alpha _{1}+\imath \frac{1}{\sqrt{3}}%
\sinh \varepsilon (\alpha _{1}+2\alpha _{2}),  \label{al1} \\
\tilde{\alpha}_{2} &=&\cosh \varepsilon \alpha _{2}-\imath \frac{1}{\sqrt{3}}%
\sinh \varepsilon (2\alpha _{1}+\alpha _{2}).  \label{al2}
\end{eqnarray}%
Different solutions to (\ref{cc}) may of course be found. With the
representation 
\begin{equation}
\sigma _{1}=\left( 
\begin{array}{rr}
-1 & 0 \\ 
1 & 1%
\end{array}%
\right) ,\sigma _{2}=\left( 
\begin{array}{rr}
1 & 1 \\ 
0 & -1%
\end{array}%
\right) ,\sigma =\sigma _{1}\sigma _{2}=\left( 
\begin{array}{rr}
-1 & -1 \\ 
1 & 0%
\end{array}%
\right) ,w_{0}=\sigma _{2}\sigma =\left( 
\begin{array}{rr}
0 & -1 \\ 
-1 & 0%
\end{array}%
\right)  \label{sigma}
\end{equation}%
the deformed longest element results with (\ref{w}) to 
\begin{equation}
w_{0}^{\varepsilon }=\left( 
\begin{array}{cc}
-\frac{2\imath }{\sqrt{3}}\sinh (2\varepsilon ) & -\cosh (2\varepsilon )-%
\frac{\imath }{\sqrt{3}}\sinh (2\varepsilon ) \\ 
-\cosh (2\varepsilon )+\frac{\imath }{\sqrt{3}}\sinh (2\varepsilon ) & \frac{%
2\imath }{\sqrt{3}}\sinh (2\varepsilon )%
\end{array}%
\right) .
\end{equation}%
It is easy to verify that (\ref{es}) is satisfied, namely $%
w_{0}^{\varepsilon }:\tilde{\alpha}_{1}\mapsto -\tilde{\alpha}_{2}^{\ast },%
\tilde{\alpha}_{2}\mapsto -\tilde{\alpha}_{1}^{\ast }.$

\subparagraph{$\mathcal{PT}$-symmetrically deformed Coxeter group factors}

Alternatively we may use the Ansatz (\ref{Ansatz}) 
\begin{equation}
\theta _{\varepsilon }=r_{0}(\varepsilon )\mathbb{I}+\imath
r_{1}(\varepsilon )(\sigma -\sigma ^{2})
\end{equation}%
where $\sigma $ is given in (\ref{sigma}). The constraint $\det \theta
_{\varepsilon }=1$, (\ref{e1}) with $s_{n}=n$ and $h=3$ yields $%
r_{0}^{2}-3r_{1}^{2}=1$ with solutions $r_{0}=\cosh \varepsilon ,r_{1}=-1/%
\sqrt{3}\sinh \varepsilon $. There are no further constraints resulting from
(\ref{e2}) as with $\vartheta =\{(e^{\imath \pi /3},e^{-\imath \pi
/3}),e^{\imath \pi 2/3},e^{-\imath \pi 2/3})\}$ it is trivially satisfied
when $r_{0}^{2}-3r_{1}^{2}=1$. Therefore we have 
\begin{equation}
\theta _{\varepsilon }=\cosh \varepsilon \mathbb{I}-\imath \frac{1}{\sqrt{3}}%
\sinh \varepsilon \left( \sigma -\sigma ^{2}\right) .
\end{equation}%
With (\ref{roots}) we obtain from this exactly the roots in (\ref{al1}) and (%
\ref{al2}), thus confirming the expectations announced at the beginning of
this subsection.

Note that in this case the constraint even holds for the individual Weyl
reflections, i.e. $\sigma _{1}\theta _{\varepsilon }=\left( \theta
_{\varepsilon }\sigma _{1}\right) ^{\ast }$ and $\sigma _{2}\theta
_{\varepsilon }=\left( \theta _{\varepsilon }\sigma _{2}\right) ^{\ast }$ as 
$\sigma _{1}=\sigma _{-}$ and $\sigma _{2}=\sigma _{+}$. This means we can
view this deformation in an alternative way as deformations across every
hyperplane in the $A_{2}$-root system. The latter was the constraint imposed
in \cite{FZ}, which explains that (\ref{al1}) and (\ref{al2}) are precisely
the deformations constructed therein.

The remaining positive nonsimple root is simply $\tilde{\alpha}_{1}+\tilde{%
\alpha}_{2}$ due to the fact that $\sigma _{\varepsilon }=\sigma $.

\noindent\textbf{{$\tilde{\Delta}(\varepsilon )$ for\ $A_{3}$\label{deltaa3}}%
}

\noindent Now the Coxeter number is even, such that according to the
reasoning after equation (\ref{w}) we expect to obtain two different types
of deformed root systems from the two different methods of construction.

\subparagraph{$\mathcal{PT}$-symmetrically deformed Coxeter group factors}

The Ansatz (\ref{Ansatz}) reads now 
\begin{equation}
\theta _{\varepsilon }=r_{0}\mathbb{I}+r_{2}\sigma ^{2}+\imath r_{1}\left(
\sigma -\sigma ^{3}\right) =\left( 
\begin{array}{ccc}
r_{0}-\imath r_{1} & -2\imath r_{1} & -\imath r_{1}-r_{2} \\ 
2\imath r_{1} & r_{0}-r_{2}+2\imath r_{1} & 2\imath r_{1} \\ 
-\imath r_{1}-r_{2} & -2\imath r_{1} & r_{0}-\imath r_{1}%
\end{array}%
\right)  \label{a3}
\end{equation}%
where we represent 
\begin{eqnarray}
\sigma _{1} &=&\left( 
\begin{array}{rrr}
-1 & 0 & 0 \\ 
1 & 1 & 0 \\ 
0 & 0 & 1%
\end{array}%
\right) ,\sigma _{2}=\sigma _{+}=\left( 
\begin{array}{rrr}
1 & 1 & 0 \\ 
0 & -1 & 0 \\ 
0 & 1 & 1%
\end{array}%
\right) ,\sigma _{3}=\left( 
\begin{array}{rrr}
1 & 0 & 0 \\ 
0 & 1 & 1 \\ 
0 & 0 & -1%
\end{array}%
\right) , \\
\sigma _{-} &=&\sigma _{1}\sigma _{3}=\left( 
\begin{array}{rrr}
-1 & 0 & 0 \\ 
1 & 1 & 1 \\ 
0 & 0 & -1%
\end{array}%
\right) ,\sigma =\left( 
\begin{array}{rrr}
-1 & -1 & 0 \\ 
1 & 1 & 1 \\ 
0 & -1 & -1%
\end{array}%
\right) ,\vartheta =\left( 
\begin{array}{ccc}
1 & -1 & 1 \\ 
-(1+\imath ) & 0 & \imath -1 \\ 
1 & 1 & 1%
\end{array}%
\right) .
\end{eqnarray}%
The constraints (\ref{e1}) and (\ref{e2}) yield 
\begin{eqnarray}
\left( r_{0}+r_{2}\right) \left[ \left( r_{0}+r_{2}\right) ^{2}-4r_{1}^{2}%
\right] &=&1,  \label{bv1} \\
r_{0}-r_{2}+2r_{1} &=&\left( r_{0}-r_{2}+2r_{1}\right) \left(
r_{0}+r_{2}\right) , \\
\left( r_{0}+r_{2}\right) &=&\left( r_{0}-r_{2}\right) ^{2}-4r_{1}^{2},
\label{bv2}
\end{eqnarray}%
where we used $s_{n}=n$ and $h=4$ to derive (\ref{bv1}). Equations (\ref{bv1}%
)-(\ref{bv2}) are solved for instance by 
\begin{equation}
r_{0}(\varepsilon )=\cosh \varepsilon ,\quad r_{1}(\varepsilon )=\pm \sqrt{%
\cosh ^{2}\varepsilon -\cosh \varepsilon }\quad \text{and\quad }%
r_{2}(\varepsilon )=1-\cosh \varepsilon .  \label{a3sol}
\end{equation}%
The three simple deformed roots are therefore 
\begin{eqnarray}
\tilde{\alpha}_{1} &=&\cosh \varepsilon \alpha _{1}+(\cosh \varepsilon
-1)\alpha _{3}-\imath \sqrt{2}\sqrt{\cosh \varepsilon }\sinh \left( \frac{%
\varepsilon }{2}\right) \left( \alpha _{1}+2\alpha _{2}+\alpha _{3}\right) ,
\\
\tilde{\alpha}_{2} &=&(2\cosh \varepsilon -1)\alpha _{2}+2\imath \sqrt{2}%
\sqrt{\cosh \varepsilon }\sinh \left( \frac{\varepsilon }{2}\right) \left(
\alpha _{1}+\alpha _{2}+\alpha _{3}\right) , \\
\tilde{\alpha}_{3} &=&\cosh \varepsilon \alpha _{3}+(\cosh \varepsilon
-1)\alpha _{1}-\imath \sqrt{2}\sqrt{\cosh \varepsilon }\sinh \left( \frac{%
\varepsilon }{2}\right) \left( \alpha _{1}+2\alpha _{2}+\alpha _{3}\right) .
\end{eqnarray}%
Making use of (\ref{ss}) the three remaining positive nonsimple roots are $%
\tilde{\alpha}_{4}:=\tilde{\alpha}_{1}+\tilde{\alpha}_{2}$, $\tilde{\alpha}%
_{5}:=\tilde{\alpha}_{2}+\tilde{\alpha}_{3}$ and $\tilde{\alpha}_{6}:=\tilde{%
\alpha}_{1}+\tilde{\alpha}_{2}+\tilde{\alpha}_{3}$.

\subparagraph{\noindent $\mathcal{CT}$-symmetrically deformed longest element%
}

We obtain an additional non-equivalent solution when $\left[ \sigma ,\theta
_{\varepsilon }\right] \neq 0$ by solving (\ref{w}). For $A_{3}$ we read off
from (\ref{case}) that $\bar{1}=3$, $\bar{2}=2$, such that (\ref{th}) leads
to the deformation matrix 
\begin{equation}
\theta _{\varepsilon }=\left( 
\begin{array}{ccc}
\theta _{11} & \theta _{12} & \theta _{13} \\ 
\theta _{21} & \theta _{22}=\theta _{22}^{\ast } & \theta _{21}^{\ast } \\ 
\theta _{13}^{\ast } & \theta _{12}^{\ast } & \theta _{11}^{\ast }%
\end{array}
\right) .
\end{equation}
Substituting this into (\ref{constw}) yields a set of constraining
equations. Assuming $\theta _{12}$ to vanish they simplify to 
\begin{eqnarray}
\theta _{22} &=&\left\vert \theta _{11}\right\vert ^{2}-\left\vert \theta
_{13}\right\vert ^{2},\quad \theta _{22}^{2}=1,\quad \left\vert \theta
_{11}\right\vert ^{2}-\theta _{13}^{2}=1,\quad  \label{a31} \\
\theta _{11}\theta _{21}^{\ast } &=&\theta _{21}(\theta _{22}+\theta
_{13}^{\ast }),\quad \theta _{11}\func{Re}\theta _{13}=0.  \label{a22}
\end{eqnarray}
Making now only the one further assumption that $\theta _{11}=\cosh
\varepsilon $ all remaining entries are fixed by (\ref{a31}) and (\ref{a22}%
). We obtain 
\begin{equation}
\theta _{\varepsilon }=\left( 
\begin{array}{ccc}
\cosh \varepsilon & 0 & \imath \sinh \varepsilon \\ 
(-\sinh ^{2}\frac{\varepsilon }{2}+\frac{\imath }{2}\sinh \varepsilon )~ & 1
& ~(-\sinh ^{2}\frac{\varepsilon }{2}-\frac{\imath }{2}\sinh \varepsilon )
\\ 
-\imath \sinh \varepsilon & 0 & \cosh \varepsilon%
\end{array}
\right) .  \label{ta3}
\end{equation}
It is easily verified that the corresponding roots have the desired
behaviour under the $\mathcal{CT}$-transformation, namely $\tilde{w}_{0}(%
\tilde{\alpha}_{1})=-\tilde{\alpha}_{3}$, $\tilde{w}_{0}(\tilde{\alpha}%
_{2})=-\tilde{\alpha}_{2}$. This solution does not correspond to a
deformation of $\sigma _{\pm }$ as now $\theta _{\varepsilon }^{\ast }\sigma
_{\pm }\neq \sigma _{\pm }\theta _{\varepsilon }$.

In this case the nonsimple roots can not be constructed from a simple
analogy to the undeformed case as $\sigma _{\varepsilon }\neq \sigma $.
Instead we have to act successively with $\sigma _{\varepsilon }$ on the
simple deformed roots. In this way the set of all positive deformed roots
results to 
\begin{eqnarray}
\tilde{\alpha}_{1} &=&\cosh \varepsilon \alpha _{1}+\imath \sinh \varepsilon
\alpha _{3}, \\
\tilde{\alpha}_{2} &=&\alpha _{2}-\sinh ^{2}\frac{\varepsilon }{2}(\alpha
_{1}+\alpha _{3})+\frac{\imath }{2}\sinh \varepsilon (\alpha _{1}-\alpha
_{3}), \\
\tilde{\alpha}_{3} &=&\cosh \varepsilon \alpha _{3}-\imath \sinh \varepsilon
\alpha _{1}, \\
\tilde{\alpha}_{4} &=&\cosh \varepsilon (\alpha _{1}+\alpha _{2})-\imath
\sinh \varepsilon (\alpha _{2}+\alpha _{3}), \\
\tilde{\alpha}_{5} &=&\cosh \varepsilon (\alpha _{2}+\alpha _{3})+\imath
\sinh \varepsilon (\alpha _{1}+\alpha _{2}), \\
\tilde{\alpha}_{6} &=&\cosh \varepsilon \alpha _{2}+\cosh ^{2}\frac{%
\varepsilon }{2}(\alpha _{1}+\alpha _{3})+\frac{\imath }{2}\sinh \varepsilon
(\alpha _{3}-\alpha _{1}).
\end{eqnarray}%
Notice that the nonsimple roots no are no longer just simple roots added
together.

\noindent\textbf{{$\tilde{\Delta}(\varepsilon )$ for\ $A_{4}$}}

\noindent Using again the Ansatz (\ref{Ansatz}) reads now 
\begin{equation}
\theta _{\varepsilon }=r_{0}(\varepsilon )\mathbb{I}+\imath
r_{1}(\varepsilon )(\sigma -\sigma ^{4})+\imath r_{2}(\varepsilon )(\sigma
^{2}-\sigma ^{3}),
\end{equation}%
with the representation 
\begin{eqnarray}
\sigma _{1} &=&\left( 
\begin{array}{rrrr}
-1 & 0 & 0 & 0 \\ 
1 & 1 & 0 & 0 \\ 
0 & 0 & 1 & 0 \\ 
0 & 0 & 0 & 1%
\end{array}%
\right) ,\sigma _{2}=\left( 
\begin{array}{rrrr}
1 & 1 & 0 & 0 \\ 
0 & -1 & 0 & 0 \\ 
0 & 1 & 1 & 0 \\ 
0 & 0 & 0 & 1%
\end{array}%
\right) ,\sigma _{3}=\left( 
\begin{array}{rrrr}
1 & 0 & 0 & 0 \\ 
0 & 1 & 1 & 0 \\ 
0 & 0 & -1 & 0 \\ 
0 & 0 & 1 & 1%
\end{array}%
\right) ,\sigma _{4}=\left( 
\begin{array}{rrrr}
1 & 0 & 0 & 0 \\ 
0 & 1 & 0 & 0 \\ 
0 & 0 & 1 & 1 \\ 
0 & 0 & 0 & -1%
\end{array}%
\right) ,~~\quad \\
\sigma &=&\sigma _{1}\sigma _{3}\sigma _{2}\sigma _{4}=\left( 
\begin{array}{rrrr}
-1 & -1 & 0 & 0 \\ 
1 & 1 & 1 & 1 \\ 
0 & -1 & -1 & -1 \\ 
0 & 1 & 1 & 0%
\end{array}%
\right) ,\quad \omega =\sigma _{2}\sigma _{4}\sigma ^{2}=\left( 
\begin{array}{rrrr}
0 & 0 & 0 & -1 \\ 
0 & 0 & -1 & 0 \\ 
0 & -1 & 0 & 0 \\ 
-1 & 0 & 0 & 0%
\end{array}%
\right) .
\end{eqnarray}%
In this case the constraints (\ref{e1}) and (\ref{e2}) yield 
\begin{eqnarray}
r_{0}^{4}-5r_{0}^{2}(r_{1}^{2}+r_{2}^{2})+5(r_{2}^{2}+r_{2}r_{1}-r_{1}^{2})^{2} &=&1,
\label{269} \\
2r_{0}^{2}+\left( -5+\sqrt{5}\right) r_{1}^{2}-\left( 5+\sqrt{5}\right)
r_{2}^{2}+4\sqrt{5}r_{1}r_{2}-2 &=&0, \\
2r_{0}+\sqrt{2\left( 5+\sqrt{5}\right) }r_{1}+\sqrt{10-2\sqrt{5}}r_{2} &\neq
&0,
\end{eqnarray}%
where we used $s_{n}=n$ and $h=5$ to obtain (\ref{269}). These equations are
solved for instance by 
\begin{equation}
r_{0}(\varepsilon )=\cosh \varepsilon ,\quad r_{1}(\varepsilon )=\kappa
_{-}\sinh \varepsilon ,\quad r_{2}(\varepsilon )=\kappa _{+}\sinh
\varepsilon ,  \label{rrr}
\end{equation}%
with $\kappa _{\pm }=\frac{1}{5}\sqrt{5\pm 2\sqrt{5}}$. The deformation
matrix results to 
\begin{equation}
\theta _{\varepsilon }=\left( 
\begin{array}{rrrr}
r_{0}-\imath r_{1} & -2\imath r_{1} & -\imath r_{1}-\imath r_{2} & -2\imath
r_{2} \\ 
2\imath r_{1} & r_{0}+2\imath r_{1}+\imath r_{2} & 2\imath r_{1}+2\imath
r_{2} & \imath r_{1}+\imath r_{2} \\ 
-\imath r_{1}-\imath r_{2} & -2\imath r_{1}-2\imath r_{2} & r_{0}-2\imath
r_{1}-\imath r_{2} & -2\imath r_{1} \\ 
2\imath r_{2} & \imath r_{1}+\imath r_{2} & 2\imath r_{1} & r_{0}+\imath
r_{1}%
\end{array}%
\right) ,
\end{equation}%
with all entries specified in (\ref{rrr}). Notice that in this case we also
obtain the deformation of the longest element $w_{0}\theta _{\varepsilon
}=\left( \theta _{\varepsilon }w_{0}\right) ^{\ast }$.

\noindent\textbf{{$\tilde{\Delta}(\varepsilon )$ for\ $A_{5}$-$A_{9}$}}

\noindent Having been very explicit in our previous examples, it suffices to
simply list the solutions for the $r_{i}$ in order to illustrate the working
of the Ansatz (\ref{Ansatz}) in the following. We find 
\begin{eqnarray}
A_{5} &:&\quad r_{1}=-r_{2}=\pm \frac{1}{\sqrt{3}}\sqrt{r_{0}^{2}-r_{0}}%
,\quad r_{3}=r_{0}-1, \\
A_{6} &:&\quad r_{1}=r_{2}=-r_{3}=1/\sqrt{7}\sqrt{r_{0}^{2}-1}, \\
A_{7} &:&\quad r_{1}=r_{3}=0,\quad r_{2}=\pm \sqrt{r_{0}^{2}-r_{0}},\quad
r_{4}=r_{0}-1, \\
A_{8} &:&\quad r_{1}=-r_{2}=-\frac{1}{3}r_{3}=r_{4}=-\frac{\sqrt{r_{0}^{2}-1}%
}{3\sqrt{3}}, \\
A_{9} &:&\quad r_{1}=-r_{4}=-\kappa _{-},\quad r_{2}=-r_{3}=-\kappa
_{+},\quad r_{5}=r_{0}-1.
\end{eqnarray}%
In all cases $r_{0}=\cosh \varepsilon $ will guarantee that also the last
constraint in (\ref{const}) is satisfied. Based on these data one may try to
conjecture closed formulae for the entire $A$-series.

\noindent\textbf{{$\tilde{\Delta}(\varepsilon )$ for\ $A_{4n-1}$}}

\noindent For the $A_{4n-1}$-subseries we succeeded to conjecture a closed
formula. Setting in (\ref{Ansatz}) all $r_{k}=0$, except for $k=0,n,2n$, the
determinant in (\ref{e1}) takes on the simple form 
\begin{equation}
\det \theta _{\varepsilon }=\left( r_{0}+r_{2n}\right) ^{2n-1}\left(
r_{0}-4r_{n}^{2}-2r_{0}r_{2n}+r_{2n}^{2}\right) ^{n},
\end{equation}%
which equals one for $r_{2n}=1-r_{0}$ and $r_{n}=\pm \sqrt{r_{0}^{2}-r_{0}}$%
. We have verified up to rank 11 that for these values the expression 
\begin{equation}
\theta _{\varepsilon }=r_{0}\mathbb{I}+r_{2n}\sigma ^{2n}+\imath r_{n}\left(
\sigma ^{n}-\sigma ^{-n}\right) ,
\end{equation}%
for the deformation matrix also satisfies the first and fourth constraint in
(\ref{const}). Once again $r_{0}=\cosh \varepsilon $ is a useful choice to
guarantee the validity of the last constraint in (\ref{const}).

\subsubsection{$\tilde{\Delta}(\protect\varepsilon )$ for\ $B_{\ell }$\label%
{bbb}}

Our convention for the labelling of the roots is to denote the short simple
root by $\alpha _{\ell }$.

\paragraph{$\tilde{\Delta}(\protect\varepsilon )$ for\ $B_{2}$}

In this case the Ansatz (\ref{Ansatz}) reads 
\begin{equation}
\theta _{\varepsilon }=r_{0}\mathbb{I}+r_{2}\sigma ^{2}+\imath r_{1}\left(
\sigma -\sigma ^{-1}\right) .
\end{equation}%
The first four constraints in (\ref{const}) are satisfied for $%
r_{0}=r_{2}\pm \sqrt{1+4r_{1}^{2}}$, which in turn is conveniently solved
for $r_{0}=\cosh \varepsilon $, $r_{2}=0$ and $r_{1}=1/2\sinh \varepsilon $,
such that simple roots and simple deformed roots are related according to (%
\ref{roots}) by 
\begin{equation}
\theta _{\varepsilon }=\left( 
\begin{array}{cc}
\cosh \varepsilon -\imath \sinh \varepsilon & -2\imath \sinh \varepsilon \\ 
\imath \sinh \varepsilon & \cosh \varepsilon +\imath \sinh \varepsilon%
\end{array}%
\right) .
\end{equation}%
This solution coincides with the one reported in \cite{Assis:2009gt}.

\paragraph{$\tilde{\Delta}(\protect\varepsilon )$ for\ $B_{3}$}

In this case the Ansatz (\ref{Ansatz}) 
\begin{equation}
\theta _{\varepsilon }=r_{0}\mathbb{I}+r_{3}\sigma ^{3}+\imath r_{1}\left(
\sigma -\sigma ^{-1}\right) +\imath r_{2}\left( \sigma ^{2}-\sigma
^{-2}\right) ,
\end{equation}%
is only solving the first four constraints in (\ref{const}) when $%
r_{0}=r_{3}-1$ and $r_{1}=-r_{2}$, which however corresponds to a trivial
real solution with $(\theta _{\varepsilon })_{ii}=-1$ for $i=1,2,3$.

\paragraph{$\tilde{\Delta}(\protect\varepsilon )$ for\ $B_{4}$}

In this case the Ansatz (\ref{Ansatz}) 
\begin{equation}
\theta _{\varepsilon }=r_{0}\mathbb{I}+r_{4}\sigma ^{4}+\imath r_{1}\left(
\sigma -\sigma ^{-1}\right) +\imath r_{2}\left( \sigma ^{2}-\sigma
^{-2}\right) +\imath r_{3}\left( \sigma ^{3}-\sigma ^{-3}\right) ,
\end{equation}
is solving the first four constraints in (\ref{const}) when $r_{0}=r_{4}\pm 
\sqrt{1+4r_{2}^{2}}$ and $r_{1}=-r_{3}$. We may incorporate the last
constraint in (\ref{const}) by solving this with $r_{0}=\cosh \varepsilon $, 
$r_{4}=0$ and $r_{2}=1/2\sinh \varepsilon $, such that the deformation
matrix becomes 
\begin{equation}
\theta _{\varepsilon }=\left( 
\begin{array}{cccc}
\cosh \varepsilon & 0 & -\imath \sinh \varepsilon & -2\imath \sinh
\varepsilon \\ 
0 & \cosh \varepsilon +\imath \sinh \varepsilon & 2\imath \sinh \varepsilon
& 2\imath \sinh \varepsilon \\ 
-\imath \sinh \varepsilon & -2\imath \sinh \varepsilon & \cosh \varepsilon
-2\imath \sinh \varepsilon & -2\imath \sinh \varepsilon \\ 
\imath \sinh \varepsilon & \imath \sinh \varepsilon & \imath \sinh
\varepsilon & \cosh \varepsilon +\imath \sinh \varepsilon%
\end{array}
\right) .
\end{equation}

\paragraph{$\tilde{\Delta}(\protect\varepsilon )$ for\ $B_{2n+1}$}

Supported by the previous examples and supplemented with several more for
higher rank not presented here, we conjecture that there are no complex
solutions for our constraints in the case of odd rank in the $B$-series
based of the Ansatz (\ref{Ansatz}).

\paragraph{$\tilde{\Delta}(\protect\varepsilon )$ for\ $B_{2n}$}

Extrapolation from $B_{2}$ and $B_{4}$ we conjecture a closed formula for
the even rank in the $B$-series 
\begin{equation}
\theta _{\varepsilon }=r_{0}\mathbb{I}+\frac{\imath }{2}r_{n}\left( \sigma
^{n}-\sigma ^{-n}\right) ,  \label{theta}
\end{equation}
for the solution of the first four constraints in (\ref{const}). It is
easily seen from (\ref{e1}) that the determinant of $\theta _{\varepsilon }$
in (\ref{theta}) results to 
\begin{equation}
\det \theta _{\varepsilon }=\prod\limits_{k=1}^{n}\left[ r_{0}-2r_{n}\sin
\left( \frac{2\pi n}{4n}s_{k}\right) \right] =\left(
r_{0}^{2}-4r_{n}^{2}\right) ^{n},  \label{dt}
\end{equation}
when using the fact that $h=4n$ and $s_{k}=2k-1$. Choosing $r_{0}=\cosh
\varepsilon $ and $r_{n}=1/2\sinh \varepsilon $ will then ensure that the
last two constraints in (\ref{const}) are also satisfied. It turns out that
the remaining equations are solved automatically without any further
restrictions. We have verified this on a case-by-case basis up to rank 8.

\subsubsection{$\tilde{\Delta}(\protect\varepsilon )$ for\ $C_{\ell }$}

This case can be solved in a completely analogous way to the $B_{n}$-series.
Equation (\ref{dt}) is absolutely identical to $B_{2n}$ and we find that the
Ansatz (\ref{theta}) together with the relevant $r_{n}$ also solves the
remaining constraints, which we have verified up to rank 8. Once again we
did not find any complex solutions up to that order of the rank for $%
C_{2n+1} $-series and conjecture also in this case that they do not exist
when based on the Ansatz (\ref{theta}).

\subsubsection{$\tilde{\Delta}(\protect\varepsilon )$ for\ $D_{\ell }$}

Our convention for the labelling of the roots is depicted in figure 1.

\subparagraph{$\mathcal{PT}$-symmetrically deformed Coxeter group factors}

For the odd rank subseries, that is $D_{2n+1}$, we find a closed formula
very similar to the one for $A_{4n-1}$. This is not surprising given the
fact that these two groups are embedded into each other as $%
D_{2n+1}\hookrightarrow A_{4n-1}$. We find that the deformation matrix of
the form 
\begin{equation}
\theta _{\varepsilon }=r_{0}\mathbb{I}+r_{2n}\sigma ^{2n}+\imath r_{n}\left(
\sigma ^{n}-\sigma ^{-n}\right) ,  \label{dd1}
\end{equation}%
solves the first four constraints in (\ref{const}) with $r_{2n}=1-r_{0}$ and 
$r_{n}=\pm \sqrt{r_{0}^{2}-r_{0}}$. The choice $r_{0}=\cosh \varepsilon $
ensures the validity of last constraint in (\ref{const}).

There are no complex solutions for $D_{2n}$ based on the Ansatz (\ref{Ansatz}%
). For instance, considering the Ansatz for $D_{4}$ the constraining
equations force us to take\ $r_{1}=-r_{2}$ and $r_{3}=r_{0}-1$, which
reduces $\theta _{\varepsilon }$ to the identity matrix $\mathbb{I}$.
Similarly the constraints for the Ansatz (\ref{Ansatz}) for $D_{6}$ demand
that $r_{1}=-r_{4}$, $r_{2}=-r_{3}$ and $r_{5}=r_{0}-1$, which reduces $%
\theta _{\varepsilon }$ again to the identity matrix $\mathbb{I}$.

\subparagraph{$\mathcal{CT}$-symmetrically deformed longest element}

For the odd rank subseries we should also be able to construct an
alternative solution by solving (\ref{constw}). As a special solution valid
for the entire subseries we find

\begin{equation}
\theta _{\varepsilon }=\left( 
\begin{array}{cc}
\mathbb{I} & 0 \\ 
0 & \hat{\theta}_{\varepsilon }%
\end{array}%
\right) ,  \label{dd2}
\end{equation}%
with 
\begin{equation}
\hat{\theta}_{\varepsilon }=\left( 
\begin{array}{ccc}
1 & (-\sinh ^{2}\frac{\varepsilon }{2}-\frac{\imath }{2}\sinh \varepsilon )
& (-\sinh ^{2}\frac{\varepsilon }{2}+\frac{\imath }{2}\sinh \varepsilon ) \\ 
0 & \cosh \varepsilon & -\imath \sinh \varepsilon \\ 
0 & -\imath \sinh \varepsilon & \cosh \varepsilon%
\end{array}%
\right) .
\end{equation}%
The solutions (\ref{dd1}) and (\ref{dd2}) do not coincide

\subsubsection{$\tilde{\Delta}(\protect\varepsilon )$ for\ $E_{6}$}

Our convention for the labelling of the roots is depicted in figure 1. \ 

\subparagraph{$\mathcal{PT}$-symmetrically deformed Coxeter group factors}

As we have seen in the previous examples we have usually more parameters at
our disposal than we require to solve the constraining equations. Thus
instead of finding the most general solution we will be content here to
solve (\ref{Ansatz}) for some restricted set of values and attempt to solve
the constraints in (\ref{const}) for 
\begin{equation}
\theta _{\varepsilon }=r_{0}\mathbb{I}+\imath r_{k}\left( \sigma ^{k}-\sigma
^{-k}\right) .
\end{equation}
Considering (\ref{e1}) for this Ansatz yields 
\begin{equation}
1=\prod\limits_{n=1}^{6}\left[ r_{0}-2r_{k}\sin \left( \frac{\pi k}{6}%
s_{n}\right) \right] \qquad \text{with }s_{n}=1,4,5,7,8,11,
\end{equation}
which reduces to 
\begin{equation}
1=\left( r_{0}^{2}-3r_{k}^{2}\right) ^{3}\qquad \text{for \ }k=2,4.
\label{cv}
\end{equation}
It turns out that in both cases the solution $r_{k}=\pm 1/\sqrt{3}\sqrt{%
r_{0}^{2}-1}$ for (\ref{cv}) also solves the first three constraints in (\ref%
{const}). For the deformation matrix we then obtain for $k=2$%
\begin{equation}
\theta _{\varepsilon }=\left( 
\begin{array}{cccccc}
r_{0} & -2\imath r_{2} & 0 & -2\imath r_{2} & -2\imath r_{2} & -\imath r_{2}
\\ 
2\imath r_{2} & r_{0}+\imath r_{2} & 2\imath r_{2} & 2\imath r_{2} & 2\imath
r_{2} & 2\imath r_{2} \\ 
0 & 2\imath r_{2} & r_{0}+2\imath r_{2} & 4\imath r_{2} & 3\imath r_{2} & 
2\imath r_{2} \\ 
-2\imath r_{2} & -2\imath r_{2} & -4\imath r_{2} & r_{0}-5\imath r_{2} & 
-4\imath r_{2} & -2\imath r_{2} \\ 
2\imath r_{2} & 2\imath r_{2} & 3\imath r_{2} & 4\imath r_{2} & 
r_{0}+2\imath r_{2} & 0 \\ 
-\imath r_{2} & -2\imath r_{2} & -2\imath r_{2} & -2\imath r_{2} & 0 & r_{0}%
\end{array}
\right) ,  \label{k2}
\end{equation}
and for $k=4$ 
\begin{equation}
\theta _{\varepsilon }=\left( 
\begin{array}{cccccc}
r_{0}-\imath r_{4} & -2\imath r_{4} & -2\imath r_{4} & -2\imath r_{4} & 0 & 0
\\ 
2\imath r_{4} & r_{0}+\imath r_{4} & 2\imath r_{4} & 2\imath r_{4} & 2\imath
r_{4} & 2\imath r_{4} \\ 
2\imath r_{4} & 2\imath r_{4} & r_{0}+3\imath r_{4} & 4\imath r_{4} & 
2\imath r_{4} & 0 \\ 
-2\imath r_{4} & -2\imath r_{4} & -4\imath r_{4} & r_{0}-5\imath r_{4} & 
-4\imath r_{4} & -2\imath r_{4} \\ 
0 & 2\imath r_{4} & 2\imath r_{4} & 4\imath r_{4} & r_{0}+3\imath r_{4} & 
2\imath r_{4} \\ 
0 & -2\imath r_{4} & 0 & -2\imath r_{4} & -2\imath r_{4} & r_{0}-\imath r_{4}%
\end{array}
\right) .  \label{k4}
\end{equation}
In each case we may specify further $r_{0}=\cosh \varepsilon $ such that $%
r_{k}=1/\sqrt{3}\sinh \varepsilon $ in order to ensure also the right
limiting behaviour, i.e. the last constraint in (\ref{const}).

\subparagraph{$\mathcal{CT}$-symmetrically deformed longest element}

We obtain an additional solution by means of the construction laid out in
section \ref{secCT}. As a particular solution we find 
\begin{equation}
\theta _{\varepsilon }=\left( 
\begin{array}{cccc}
1 & 0 & 0 & 0 \\ 
0 & 1 & 0 & 0 \\ 
0 & 0 & \theta _{\varepsilon }^{A_{3}} & 0 \\ 
0 & 0 & 0 & 1%
\end{array}%
\right) ,  \label{e6def}
\end{equation}%
with $\theta _{\varepsilon }^{A_{3}}$ given in (\ref{ta3}). This means the
fact that the subsystem made from the vertices $3,4$ and $5$ is identical to 
$A_{3}$ also reflects in the solution for the deformation matrix. Clearly
this solution is different from (\ref{k2}) as well as (\ref{k4}).

\subsubsection{$\tilde{\Delta}(\protect\varepsilon )$ for\ $E_{7}$}

Our convention for the labelling of the roots is the same as for $E_{6}$ by
linking the additional root $\alpha _{7}$ to $\alpha _{6}$. There exists no
complex solution to (\ref{const}) based on the Ansatz (\ref{Ansatz}) with $%
h=18$. Together with the explicit representation for the $\sigma $ we
substitute this into the constraints (\ref{const}) and find the unique real
solution for the unknown functions $r_{0}=1+r_{5}$, $%
r_{1}=-r_{4}-r_{5}-r_{8} $, $r_{2}=-r_{4}-r_{5}-r_{7}$ and $r_{3}=-r_{6}$,
which reduced the deformation matrix to the identity operator $\theta
_{\varepsilon }=\mathbb{I}$.

\subsubsection{$\tilde{\Delta}(\protect\varepsilon )$ for\ $E_{8}$}

Our convention for the labelling of the roots is the same as for $E_{7}$ by
linking the additional root $\alpha _{8}$ to $\alpha _{7}$. Also in this
case there exists no complex solution to (\ref{const}) based on the Ansatz (%
\ref{Ansatz}) with $h=30$. When substituted into the constraints (\ref{const}%
) we find the unique solution $r_{0}=1+r_{5}$, $%
r_{1}=-r_{5}-r_{6}-r_{9}-r_{10}-r_{14}$, $%
r_{2}=-2r_{5}-r_{7}-r_{8}-2r_{10}-r_{13}$,\ $%
r_{3}=-r_{5}-r_{7}-r_{8}-r_{10}-r_{12}$ and $%
r_{4}=r_{5}-r_{6}-r_{9}+r_{10}-r_{11}$. However, this solution simply
corresponds to $\theta _{\varepsilon }=\mathbb{I}$.

\subsubsection{$\tilde{\Delta}(\protect\varepsilon )$ for\ $F_{4}$}

Our convention for the labelling of the roots is to denote the long roots by 
$\alpha _{1}$,$\alpha _{2}$ and short roots by $\alpha _{3}$,$\alpha _{4}$
with $\alpha _{i}$ linked to $\alpha _{i+1}$ for $i=1,2,3$. In the $F_{4}$%
-Ansatz (\ref{Ansatz}) 
\begin{equation}
\theta _{\varepsilon }=r_{0}\mathbb{I}+r_{6}\sigma ^{6}+\imath
\sum\limits_{k=1}^{5}r_{k}\left( \sigma ^{k}-\sigma ^{-k}\right)
\end{equation}%
we have seven unknown functions left. We find two inequivalent solutions for
the first four constraints in (\ref{const}), when specifying only two
functions, either 
\begin{equation}
r_{1}=-2r_{3}-r_{5}\pm \sqrt{(r_{0}-r_{6})^{2}-1}\qquad \text{and\qquad }%
r_{2}=-r_{4}  \label{r6}
\end{equation}%
or 
\begin{equation}
r_{1}=-2r_{3}-r_{5}\qquad \text{and\qquad }r_{2}=-r_{4}\pm \frac{1}{\sqrt{3}}%
\sqrt{(r_{0}-r_{6})^{2}-1}.  \label{r62}
\end{equation}%
This leaves five functions at our disposal, which we may choose in
accordance with the last constraint in (\ref{const}). Taking for instance $%
r_{3}=r_{4}=r_{5}=r_{6}=0$ and $r_{0}=\cosh \varepsilon $ in (\ref{r6})
yields 
\begin{equation}
\theta _{\varepsilon }=\left( 
\begin{array}{cccc}
\cosh \varepsilon -\imath \sinh \varepsilon & -2\imath \sinh \varepsilon & 
-2\imath \sinh \varepsilon & 0 \\ 
2\imath \sinh \varepsilon & \cosh \varepsilon +3\imath \sinh \varepsilon & 
4\imath \sinh \varepsilon & 2\imath \sinh \varepsilon \\ 
-\imath \sinh \varepsilon & -2\imath \sinh \varepsilon & \cosh \varepsilon
-3\imath \sinh \varepsilon & -2\imath \sinh \varepsilon \\ 
0 & \imath \sinh \varepsilon & 2\imath \sinh \varepsilon & \cosh \varepsilon
+\imath \sinh \varepsilon%
\end{array}%
\right) ,
\end{equation}%
for the deformation matrix.

\subsubsection{$\tilde{\Delta}(\protect\varepsilon )$ for\ $G_{2}$}

We label the short root by $\alpha _{1}$ and the long root by $\alpha _{2}$.
\ As mentioned, this case has been solved before \cite{FZ}, but nonetheless
we report it here for completeness and to demonstrate that it fits well into
the general framework provided here. The Ansatz (\ref{Ansatz}) with $h=6$
solves the first four constraints (\ref{const}) uniquely with $r_{3}=0$ and $%
r_{0}=\pm \sqrt{1+3(r_{1}+r_{2})^{2}}$. The choice $r_{1}=1/\sqrt{3}\sinh
\varepsilon -r_{2}$ reproduces the result of \cite{FZ}.

This completes the study of all crystallographic Coxeter groups. We will
also consider one\ noncrystallographic example.

\subsubsection{$\tilde{\Delta}(\protect\varepsilon )$ for\ $H_{3}$}

We label the long roots by $\alpha _{1}$, $\alpha _{2}$ and the short root
by $\alpha _{3}$. In this case there are no complex solutions of the type we
are seeking here. Substituting the Ansatz (\ref{Ansatz}) with $h=6$ into the
constraints (\ref{const}) leads to the unique solution $r_{0}=1$, $r_{5}=0$
and $r_{1}+r_{4}=-\phi (r_{2}+r_{3})$ with $\phi $ being the golden ratio $%
\phi =(1+\sqrt{5})/2$ appearing in the $H_{3}$-Cartan matrix. However, this
solution simply corresponds to $\theta _{\varepsilon }=\mathbb{I}$.

\subsubsection{Solutions from folding}

One deficiency of the above constructions is that in some cases they do not
lead to any complex solution for $\tilde{\Delta}$. However, we demonstrate
now that in these cases one may still construct higher dimensional solutions
by means of the so-called folding procedure, see e.g. \cite%
{DIO,Sasaki:1991br,PratikKhastgir:1995ur,Fring:2005am,Fring:2005va}. This
construction makes use of the fact that some root systems are embedded into
larger ones. Identifying roots which are related by the involution (\ref%
{case}), one obtains a root system associated to a different type of Coxeter
group. At the same time we may use the folding procedure for consistency
checks.

\paragraph{$B_{n}\hookrightarrow A_{2n}$}

We showed that there exist no complex deformations for the $B_{2n-1}$-series
based on the Ansatz (\ref{Ansatz}). However, making use of the embedding $%
B_{n}\hookrightarrow A_{2n}$ we demonstrate now that one can construct
higher dimensional solutions from the reduction of $A_{4n-2}$ to $B_{2n-1}$.
We illustrate this in detail for the particular case $B_{3}\hookrightarrow
A_{6}$. Starting with the solution to the constraints (\ref{const}) for $%
A_{6}$-deformation matrix 
\begin{equation}
\theta _{\varepsilon }=r_{0}\mathbb{I}+\imath r_{1}\left( \sigma -\sigma
^{-1}\right) +\imath r_{2}\left( \sigma ^{2}-\sigma ^{-2}\right) +\imath
r_{3}\left( \sigma ^{3}-\sigma ^{-3}\right) ,
\end{equation}%
with $r_{1}=r_{2}=-r_{3}=1/\sqrt{7}\cosh \varepsilon $, we employ the
explicit form for $\sigma $ to obtain the simple deformed $A_{6}$-roots from
(\ref{roots}) 
\begin{eqnarray}
\tilde{\alpha}_{1} &=&\cosh \varepsilon \alpha _{1}-\imath /\sqrt{7}\sinh
\varepsilon (\alpha _{1}+2\alpha _{2}+2\alpha _{3}+2\alpha _{4}-2\alpha
_{6}), \\
\tilde{\alpha}_{2} &=&\cosh \varepsilon \alpha _{2}+\imath /\sqrt{7}\sinh
\varepsilon (2\alpha _{1}+3\alpha _{2}+4\alpha _{3}+2\alpha _{4}), \\
\tilde{\alpha}_{3} &=&\cosh \varepsilon \alpha _{3}-\imath /\sqrt{7}\sinh
\varepsilon (2\alpha _{1}+4\alpha _{2}+3\alpha _{3}+2\alpha _{4}+2\alpha
_{5}+2\alpha _{6}), \\
\tilde{\alpha}_{4} &=&\cosh \varepsilon \alpha _{4}+\imath /\sqrt{7}\sinh
\varepsilon (2\alpha _{1}+2\alpha _{2}+2\alpha _{3}+3\alpha _{4}+4\alpha
_{5}+2\alpha _{6}), \\
\tilde{\alpha}_{5} &=&\cosh \varepsilon \alpha _{5}-\imath /\sqrt{7}\sinh
\varepsilon (2\alpha _{3}+4\alpha _{4}+3\alpha _{5}+2\alpha _{6}), \\
\tilde{\alpha}_{6} &=&\cosh \varepsilon \alpha _{6}-\imath /\sqrt{7}\sinh
\varepsilon (2\alpha _{1}-2\alpha _{3}-2\alpha _{4}-2\alpha _{5}-\alpha
_{6}).
\end{eqnarray}%
Following the folding procedure we can now define deformed simple $B_{3}$%
-roots as 
\begin{eqnarray}
\tilde{\beta}_{1} &=&\tilde{\alpha}_{1}+\tilde{\alpha}_{6}=\cosh \varepsilon
(\alpha _{1}+\alpha _{6})-\imath /\sqrt{7}\sinh \varepsilon \lbrack 3(\alpha
_{1}-\alpha _{6})+2(\alpha _{2}-\alpha _{5})], \\
\tilde{\beta}_{2} &=&\tilde{\alpha}_{2}+\tilde{\alpha}_{5}=\cosh \varepsilon
(\alpha _{2}+\alpha _{5})+\imath /\sqrt{7}\sinh \varepsilon \lbrack 2(\alpha
_{1}-\alpha _{6}+\alpha _{3}-\alpha _{4})+\alpha _{2}-\alpha _{5}],~~~~~~~ \\
\tilde{\beta}_{3} &=&\tilde{\alpha}_{3}+\tilde{\alpha}_{4}=\cosh \varepsilon
(\alpha _{1}+\alpha _{6})-\imath /\sqrt{7}\sinh \varepsilon \lbrack 2(\alpha
_{2}-\alpha _{5})+\alpha _{3}-\alpha _{4}].
\end{eqnarray}%
These roots reproduce the $B_{3}$-Cartan matrix, but it is not possible to
express the imaginary part in terms of the undeformed $B_{3}$-roots$.$ As
expected from section \ref{bbb}, it is therefore impossible to find a three
dimensional deformation matrix of the type (\ref{roots}). When identifying
the undeformed $A_{6}$-roots related by the involution (\ref{case})
according to $\alpha _{1}\leftrightarrow \alpha _{6}$, $\alpha
_{2}\leftrightarrow \alpha _{5}$ and $\alpha _{3}\leftrightarrow \alpha _{4}$%
, the deformed $B_{3}$-roots will all become real.

\paragraph{$F_{4}\hookrightarrow E_{6}$}

Having found some new solutions for a case which could not be solved
previously, let us see next how some solutions we have found are related to
each other through the folding procedure. In analogy to the undeformed case
we may define the deformed $F_{4}$-roots in terms of the deformed $E_{6}$%
-roots as 
\begin{equation}
\tilde{\beta}_{1}^{F_{4}}=\tilde{\alpha}_{1}^{E_{6}}+\tilde{\alpha}%
_{6}^{E_{6}},\quad \quad \tilde{\beta}_{2}^{F_{4}}=\tilde{\alpha}%
_{3}^{E_{6}}+\tilde{\alpha}_{5}^{E_{6}},\quad \quad \tilde{\beta}%
_{3}^{F_{4}}=\tilde{\alpha}_{4}^{E_{6}}\quad \text{and\quad }\tilde{\beta}%
_{4}^{F_{4}}=\tilde{\alpha}_{3}^{E_{6}}.  \label{fold}
\end{equation}
This means the $F_{4}$-deformation matrix is constructed as 
\begin{equation}
\theta _{\varepsilon }^{F_{4}}=\left( 
\begin{array}{cccc}
\frac{\theta _{11}^{E_{6}}+\theta _{61}^{E_{6}}+\theta _{16}^{E_{6}}+\theta
_{66}^{E_{6}}}{2} & \frac{\theta _{13}^{E_{6}}+\theta _{63}^{E_{6}}+\theta
_{15}^{E_{6}}+\theta _{65}^{E_{6}}}{2} & \theta _{14}^{E_{6}}+\theta
_{64}^{E_{6}} & \theta _{12}^{E_{6}}+\theta _{62}^{E_{6}} \\ 
\frac{\theta _{31}^{E_{6}}+\theta _{51}^{E_{6}}+\theta _{36}^{E_{6}}+\theta
_{56}^{E_{6}}}{2} & \frac{\theta _{33}^{E_{6}}+\theta _{53}^{E_{6}}+\theta
_{35}^{E_{6}}+\theta _{55}^{E_{6}}}{2} & \theta _{34}^{E_{6}}+\theta
_{54}^{E_{6}} & \theta _{32}^{E_{6}}+\theta _{52}^{E_{6}} \\ 
\frac{\theta _{41}^{E_{6}}+\theta _{46}^{E_{6}}}{2} & \frac{\theta
_{43}^{E_{6}}+\theta _{45}^{E_{6}}}{2} & \theta _{44}^{E_{6}} & \theta
_{42}^{E_{6}} \\ 
\frac{\theta _{21}^{E_{6}}+\theta _{26}^{E_{6}}}{2} & \frac{\theta
_{23}^{E_{6}}+\theta _{25}^{E_{6}}}{2} & \theta _{24}^{E_{6}} & \theta
_{22}^{E_{6}}%
\end{array}
\right) .
\end{equation}
In this reduction the two inequivalent deformed $E_{6}$-root systems (\ref%
{k2}) and (\ref{k4}) produce the same solution for $F_{4}$%
\begin{equation}
\theta _{\varepsilon }^{F_{4}}=\left( 
\begin{array}{cccc}
r_{0}-\imath r_{k} & -2\imath r_{k} & -4\imath r_{k} & -4\imath r_{k} \\ 
2\imath r_{k} & r_{0}+5\imath r_{k} & 8\imath r_{k} & 4\imath r_{k} \\ 
-2\imath r_{k} & -4\imath r_{k} & r_{0}-5\imath r_{k} & -2\imath r_{k} \\ 
2\imath r_{k} & 2\imath r_{k} & 2\imath r_{k} & r_{0}+\imath r_{k}%
\end{array}
\right) .
\end{equation}
This solution corresponds to a special solution we found in the context of $%
F_{4}$, namely (\ref{r62}) with $r_{4}=r_{6}=0$.

Using the same identification between the $F_{4}$ and $E_{6}$ roots as in (%
\ref{fold}), we obtain from the solution based on the deformation of the
longest element (\ref{e6def}) 
\begin{eqnarray}
\tilde{\beta}_{1}^{F_{4}} &=&\alpha _{1}^{E_{6}}+\alpha _{6}^{E_{6}}, \\
\tilde{\beta}_{2}^{F_{4}} &=&(\cosh \varepsilon -\imath \sinh \varepsilon
)\alpha _{3}^{E_{6}}+(\cosh \varepsilon +\imath \sinh \varepsilon )\alpha
_{5}^{E_{6}}, \\
\tilde{\beta}_{3}^{F_{4}} &=&\frac{1}{2}(1-\cosh \varepsilon +\imath \sinh
\varepsilon )\alpha _{3}^{E_{6}}+\alpha _{4}^{E_{6}}+\frac{1}{2}(1-\cosh
\varepsilon -\imath \sinh \varepsilon )\alpha _{5}^{E_{6}} \\
\tilde{\beta}_{4}^{F_{4}} &=&\cosh \varepsilon \alpha _{3}^{E_{6}}+\imath
\sinh \varepsilon \alpha _{5}^{E_{6}}.
\end{eqnarray}%
These roots reproduce the $F_{4}$-Cartan matrix, but it is not possible to
express them in terms of the undeformed $F_{4}$-roots$.$ This reflects the
fact that the longest elements acts trivially in this case and therefore
also no nontrivial deformation of this involution exists.

\section{Antilinear deformations of Calogero models}

We have constructed a deformation map $\delta $ which replaces each root $%
\alpha $ by a deformed counterpart $\tilde{\alpha}$ as specified above. We
will now employ this construction in the context of a concrete physical
model and replace the set of $n$-dynamical variables $q=\{q_{1},\ldots
,q_{n}\}$ and their conjugate momenta $p=\{p_{1},\ldots ,p_{n}\}$ by means
of this deformation map $\delta :(q,p)\rightarrow (\tilde{q},\tilde{p})$.

\subsection{The l=0 wavefunctions and eigenenergies in the undeformed case}

Let us first generalize Calogero's construction \cite{Cal1} for the solution
of the $l=0$ wavefunction to generic Coxeter groups $\mathcal{W}$. We
consider the generalized\footnote{%
In the sense of being not dependent on a specific representation of the
roots and a particular Coxeter group.} Calogero Hamiltonian 
\begin{equation}
\mathcal{H}_{C}(p,q)=\frac{p^{2}}{2}+\frac{\omega ^{2}}{4}\sum_{\alpha \in
\Delta ^{+}}(\alpha \cdot q)^{2}+\sum_{\alpha \in \Delta ^{+}}\frac{%
g_{\alpha }}{(\alpha \cdot q)^{2}},  \label{HC}
\end{equation}%
with\ $g_{\alpha }$ being real coupling constants, which for the time being
may be different for each positive root $\alpha \in \Delta ^{+}$ associated
to any Coxeter group $\mathcal{W}$. Generalizing \cite{Cal1} we define now
the variables 
\begin{equation}
z:=\prod\limits_{\alpha \in \Delta ^{+}}(\alpha \cdot q)\qquad \text{%
and\qquad }r^{2}:=\frac{1}{\hat{h}t_{\ell }}\sum\limits_{\alpha \in \Delta
^{+}}(\alpha \cdot q)^{2},  \label{rz}
\end{equation}%
where $\hat{h}$ denotes the dual Coxeter number and $t_{\ell }$ is the $\ell 
$-th symmetrizer of the incidence matrix $I$ defined through the relation $%
I_{ij}t_{j}=t_{i}I_{ij}$, see appendix B for some concrete values. We assume
next that the wavefunction can be separated in terms of these variables in
the form 
\begin{equation}
\psi (q)\rightarrow \psi (z,r)=z^{\kappa +1/2}\varphi (r),  \label{ansatz}
\end{equation}%
with $\kappa $ being an undetermined constant for the moment. Using this
Ansatz we try to solve the $n$-body Schr\"{o}dinger equation in position
space $\mathcal{H}_{C}\psi (q)=E\psi (q)$ with\ $p^{2}=-\sum%
\nolimits_{i=1}^{n}\partial _{q_{i}}^{2}$. Changing variables for the
Laplace operator then yields 
\begin{eqnarray}
\left\{ -\frac{1}{2}\sum\limits_{i=1}^{n}\left[ \left( \kappa ^{2}-\frac{1}{4%
}\right) \frac{1}{z^{2}}\left( \frac{\partial z}{\partial q_{i}}\right)
^{2}+\left( \kappa +\frac{1}{2}\right) \frac{1}{z}\left( \frac{\partial ^{2}z%
}{\partial q_{i}^{2}}+2\frac{\partial z}{\partial q_{i}}\frac{\partial r}{%
\partial q_{i}}\frac{\partial }{\partial r}\right) +\frac{\partial ^{2}r}{%
\partial q_{i}^{2}}\frac{\partial ^{2}}{\partial r^{2}}\right. \right. \ \ \
\ \ \ \ \  &&  \label{he} \\
\left. \left. +\left( \frac{\partial r}{\partial q_{i}}\right) ^{2}\frac{%
\partial }{\partial r}\right] +\frac{\omega ^{2}}{4}\hat{h}t_{\ell
}r^{2}+\sum_{\alpha \in \Delta ^{+}}\frac{g_{\alpha }}{(\alpha \cdot q)^{2}}%
-E\right\} \varphi (r) &=&0.  \notag
\end{eqnarray}%
Taking now $g_{\alpha }=g\alpha ^{2}/2$, i.e. having the same coupling
constant for all short and all long roots, and using the identities (\ref{21}%
)-(\ref{auch}) from appendix A this reduces to 
\begin{equation}
\left\{ -\frac{1}{2}\left[ \frac{\partial ^{2}}{\partial r^{2}}+\left[
\left( \kappa +\frac{1}{2}\right) h\ell +(\ell +1)\right] \frac{1}{r}\frac{%
\partial }{\partial r}\right] +\frac{\omega ^{2}}{4}\hat{h}t_{\ell
}r^{2}\right\} \varphi (r)=E\varphi (r).  \label{radial}
\end{equation}%
The key feature is that due to the identity (\ref{21}) the first term in (%
\ref{he}) combines with part of the potential term to 
\begin{equation}
\left[ \frac{g}{2}-\frac{1}{2}\left( \kappa ^{2}-\frac{1}{4}\right) \right]
\sum_{\alpha \in \Delta ^{+}}\frac{\alpha ^{2}}{(\alpha \cdot q)^{2}}.
\end{equation}%
This term vanishes when choosing the free parameter $\kappa $ to $\kappa
=\pm 1/2\sqrt{1+4g}$. The positive solution is the only physical acceptable
one, as we would obtain singularities in (\ref{ansatz}) and therefore a
nonnormalizable wavefunction otherwise.

The equation (\ref{radial}) is a second order differential equation which
may be solved by standard methods. Imposing as usual the physical constraint
that the wavefunction vanishes at infinity, the energy quantizes to 
\begin{equation}
E_{n}=\frac{1}{4}\left[ \left( 2+h+h\sqrt{1+4g}\right) l+8n\right] \sqrt{%
\frac{\hat{h}t_{\ell }}{2}}\omega  \label{en}
\end{equation}
with corresponding wavefunctions 
\begin{equation}
\varphi _{n}(r)=c_{n}\exp \left( -\sqrt{\frac{\hat{h}t_{\ell }}{2}}\frac{%
\omega }{2}r^{2}\right) L_{n}^{a}\left( \sqrt{\frac{\hat{h}t_{\ell }}{2}}%
\omega r^{2}\right) .  \label{phin}
\end{equation}
Here $L_{n}^{a}(x)$ denotes the generalized Laguerre polynomial, $c_{n}$ is
a normalization constant and $a=\left( 2+h+h\sqrt{1+4g}\right) l/4-1$.

A key feature of the model is that the last term in the potential in (\ref%
{HC}) becomes singular whenever $q_{i}=q_{j}$ for any $i,j\in \{1,2,\ldots
,n\}$. This means that the wavefunction is vanishing at these points and we
may encounter nontrivial phases for any two particle interchange. In fact,
as the variable $z$ defined in (\ref{rz}) is antisymmetric and $r$ is
symmetric in all variables it is easy to see that the associated particles
give rise to anyonic exchange factors 
\begin{equation}
\psi (q_{1},\ldots ,q_{i},q_{j},\ldots q_{n})=e^{\imath \pi s}\psi
(q_{1},\ldots ,q_{j},q_{i},\ldots q_{n}),\quad \text{for }1\leq i,j\leq n,
\end{equation}%
with 
\begin{equation}
s=\frac{1}{2}+\frac{1}{2}\sqrt{1+4g}.  \label{anyon}
\end{equation}%
This property of the model will change in the deformed case.

\subsection{The l=0 wavefunctions and eigenenergies in the deformed case}

Now we consider the antilinear deformation of the Calogero Hamiltonian $%
\delta :\mathcal{H}_{C}(\alpha )\rightarrow \mathcal{H}_{adC}(\tilde{\alpha})
$ 
\begin{equation}
\mathcal{H}_{adC}(p,q)=\frac{p^{2}}{2}+\frac{\omega ^{2}}{4}\sum_{\tilde{%
\alpha}\in \tilde{\Delta}^{+}}(\tilde{\alpha}\cdot q)^{2}+\sum_{\tilde{\alpha%
}\in \Delta ^{+}}\frac{g_{\tilde{\alpha}}}{(\tilde{\alpha}\cdot q)^{2}}.
\label{adC}
\end{equation}%
In analogy to the deformed case we attempt to solve this model by a similar
reparameterization as (\ref{rz}), i.e. defining the variables 
\begin{equation}
\tilde{z}:=\prod\limits_{\tilde{\alpha}\in \tilde{\Delta}^{+}}(\tilde{\alpha}%
\cdot q)\qquad \text{and\qquad }\tilde{r}^{2}:=\frac{1}{\hat{h}t_{\ell }}%
\sum\limits_{\tilde{\alpha}\in \tilde{\Delta}^{+}}(\tilde{\alpha}\cdot
q)^{2},  \label{rzt}
\end{equation}%
and separating the wavefunction as 
\begin{equation}
\psi (q)\rightarrow \psi (\tilde{z},\tilde{r})=\tilde{z}^{s}\varphi (\tilde{r%
}).  \label{an2}
\end{equation}%
As a consequence of our construction for the deformed roots for which we
demanded that inner products are preserved, we find that $\tilde{r}=r$.
Furthermore, we observe that due to this fact the relations (\ref{12}) and (%
\ref{14}) also hold when replacing $\alpha $ by $\tilde{\alpha}$ and
consequently the solution procedure for the eigenvalue equation does not
change. Therefore we obtain 
\begin{equation}
\psi (q)=\psi (\tilde{z},r)=\tilde{z}^{s}\varphi _{n}(r)  \label{pht}
\end{equation}%
as solution with $\varphi _{n}(r)$ given in (\ref{phin}) and unchanged
energy eigenvalues (\ref{en}). When generalizing the ansatz (\ref{an2}) to
take also values for $l\neq 0$ into account the energy eigenvalues will,
however, change, as was demonstrated in \cite{FZ} for $A_{2}$ and $G_{2}$.
The main difference between the deformed and undeformed case for the
solution provided here is the occurrence of the variable $\tilde{z}$ instead
of $z$. As a consequence the wavefunction (\ref{pht}) no longer vanishes
when two $q_{i}$s values coincide, which in turn is a reflection of the fact
that all singularities resulting from a two-particle exchange have been
regularized by means of the deformation. However, we still encounter
singularities in the potential when all $n$ values for the $q_{i}$s
coincide. The wavefunction vanishes in this case and we obtain nontrivial
statistics exchange factors. 

Let us see in detail for some concrete models how to obtain nontrivial
anyonic exchange factors for an $n$-particle scattering process.

\subsubsection{The deformed $A_{2}$-model}

The potential in (\ref{adC}) and the variable $\tilde{z}$ in (\ref{rzt}) are
computed from the inner products of all $3$ roots in $\tilde{\Delta}%
_{A_{2}}^{+}$ with the vector $q$. Using the standard three dimensional
representation for the simple $A_{2}$-roots $\alpha _{1}=\{1,-1,0\}$ and $%
\alpha _{2}=\{0,1,-1\}$, we find with (\ref{al1}) and (\ref{al2}) 
\begin{eqnarray}
\tilde{\alpha}_{1}\cdot q &=&q_{12}\cosh \varepsilon -\frac{\imath }{\sqrt{3}%
}(q_{13}+q_{23})\sinh \varepsilon , \\
\tilde{\alpha}_{2}\cdot q &=&q_{23}\cosh \varepsilon -\frac{\imath }{\sqrt{3}%
}(q_{21}+q_{31})\sinh \varepsilon , \\
(\tilde{\alpha}_{1}+\tilde{\alpha}_{2})\cdot q &=&q_{13}\cosh \varepsilon +%
\frac{\imath }{\sqrt{3}}(q_{12}+q_{32})\sinh \varepsilon .
\end{eqnarray}
For convenience we introduced the notation $q_{ij}:=q_{i}-q_{j}$. The new
feature of these models is that the last term in the potential (\ref{adC})
resulting from these products is no longer singular when the position of two
particles coincides. It is easy to see that the $\mathcal{PT}$-symmetry
constructed for the $\tilde{\alpha}$ may be realized alternatively in the
dual space, that is on the level of the dynamical variables 
\begin{eqnarray}
\sigma _{-}^{\varepsilon } &:&\quad \tilde{\alpha}_{1}\leftrightarrow -%
\tilde{\alpha}_{1}\text{, }\tilde{\alpha}_{2}\leftrightarrow \tilde{\alpha}%
_{1}+\tilde{\alpha}_{2}\quad \Leftrightarrow \quad q_{1}\leftrightarrow q_{2}%
\text{, }q_{3}\leftrightarrow q_{3}\text{, }\imath \rightarrow -\imath , \\
\sigma _{+}^{\varepsilon } &:&\quad \tilde{\alpha}_{2}\leftrightarrow -%
\tilde{\alpha}_{2}\text{, }\tilde{\alpha}_{1}\leftrightarrow \tilde{\alpha}%
_{1}+\tilde{\alpha}_{2}\quad \Leftrightarrow \quad q_{2}\leftrightarrow q_{3}%
\text{, }q_{1}\leftrightarrow q_{1}\text{, }\imath \rightarrow -\imath .
\end{eqnarray}
A crucial difference to the undeformed case is that $\tilde{z}$ will, unlike 
$z$, not vanish in the two particle scattering process when two $q$s
coincide. In fact in that case $\tilde{z}$ will be purely imaginary as
follows directly from the $\mathcal{PT}$-symmetry 
\begin{eqnarray}
\sigma _{-}^{\varepsilon }\tilde{z}(q_{1},q_{2},q_{3}) &=&\tilde{z}^{\ast
}(q_{2},q_{1},q_{3})=-\tilde{z}(q_{1},q_{2},q_{3})\quad \Rightarrow \quad 
\tilde{z}(q_{1},q_{1},q_{3})\in \imath \mathbb{R},  \label{sm} \\
\sigma _{+}^{\varepsilon }\tilde{z}(q_{1},q_{2},q_{3}) &=&\tilde{z}^{\ast
}(q_{1},q_{3},q_{2})=-\tilde{z}(q_{1},q_{2},q_{3})\quad \Rightarrow \quad 
\tilde{z}(q_{1},q_{3},q_{3})\in \imath \mathbb{R}.  \label{sp}
\end{eqnarray}
The remaining possibility $\tilde{z}(q_{1},q_{2},q_{1})\in \imath \mathbb{R}$
follows from the previous cases together with the cyclic property $\tilde{z}%
(q_{1},q_{2},q_{3})=\tilde{z}(q_{2},q_{3},q_{1})$, which in turn results
when combining (\ref{sm}) and (\ref{sp}). Under these circumstances a new
symmetry arises 
\begin{equation}
\alpha _{1}=0\text{, }\alpha _{2}\rightarrow -\alpha _{2}\quad
\Leftrightarrow \quad \tilde{\alpha}_{1}\rightarrow -\tilde{\alpha}_{1}\text{%
, }\tilde{\alpha}_{2}\rightarrow -\tilde{\alpha}_{2}\quad \Leftrightarrow
\quad q_{1}=q_{2}\text{, }q_{2}\leftrightarrow q_{3},
\end{equation}
leading to $\tilde{z}(q_{2},q_{2},q_{3})=-\tilde{z}(q_{3},q_{3},q_{2})$. By (%
\ref{ansatz}) this means 
\begin{equation}
\psi (q_{2},q_{2},q_{3})=e^{\imath \pi s}\psi (q_{3},q_{3},q_{2})
\end{equation}
with $s$ given in (\ref{anyon}). Hence we obtain a nontrivial exchange
factor in the three-particle scattering process when particle $1$ and $2$
have the same position and are simultaneously scattered with particle $3$.

\noindent Similarly we observe 
\begin{equation}
\alpha _{2}=0\text{, }\alpha _{1}\rightarrow -\alpha _{1}\quad
\Leftrightarrow \quad \tilde{\alpha}_{1}\rightarrow -\tilde{\alpha}_{1}\text{%
, }\tilde{\alpha}_{2}\rightarrow -\tilde{\alpha}_{2}\quad \Leftrightarrow
\quad q_{2}=q_{3}\text{, }q_{1}\leftrightarrow q_{2},
\end{equation}
leading to $\tilde{z}(q_{1},q_{2},q_{2})=-\tilde{z}(q_{2},q_{2},q_{1})$ and
therefore 
\begin{equation}
\psi (q_{1},q_{2},q_{2})=e^{\imath \pi s}\psi (q_{2},q_{2},q_{1}).
\end{equation}
Now a nontrivial exchange factor emerges in the three-particle scattering
process when particle $2$ and $3$ have the same position and are
simultaneously scattered with particle $1$. We depict various possibilities
in figure $2$.

\unitlength=0.6500000pt 
\begin{picture}(300.0,70.00)(150.00,125.00)

\put(250.00,150.00){\circle*{10.00}}
\put(295.00,150.00){\circle*{10.00}}
\put(340.00,150.00){\circle*{10.00}}

\put(245.00,165.00){$ {\small x } $}
\put(290.00,165.00){$ {\small y } $}
\put(335.00,165.00){$ {\small z } $}

\put(245.00,132.00){$ {\small q_{1}} $}
\put(290.00,132.00){$ {\small q_{2}} $}
\put(335.00,132.00){$ {\small q_{3}} $}

\put(190.00,150.00){\line(1,0){210.00}}

\put(420.00,147.00){$ = $}

\put(540.00,150.00){\circle*{10.00}}
\put(585.00,150.00){\circle*{10.00}}
\put(630.00,150.00){\circle*{10.00}}

\put(535.00,132.00){$ {\small q_{2}} $}
\put(580.00,132.00){$ {\small q_{3}} $}
\put(625.00,132.00){$ {\small q_{1}} $}

\put(535.00,165.00){$ {\small x } $}
\put(580.00,165.00){$ {\small y } $}
\put(625.00,165.00){$ {\small z } $}

\put(480.00,150.00){\line(1,0){210.00}}
\end{picture}

\unitlength=0.6500000pt 
\begin{picture}(300.0,70.00)(150.00,125.00)
\put(247.00,150.00){\circle*{10.00}}
\put(253.00,150.00){\circle*{10.00}}
\put(340.00,150.00){\circle*{10.00}}
\put(245.00,165.00){$ {\small x } $}
\put(335.00,165.00){$ {\small y } $}
\put(225.00,132.00){$ {\small q_{1}=q_{2}} $}
\put(335.00,132.00){$ {\small q_{3}} $}
\put(190.00,150.00){\line(1,0){210.00}}
\put(420.00,147.00){$ = \,\, e^{\imath \pi s} $}
\put(540.00,150.00){\circle*{10.00}}
\put(627.00,150.00){\circle*{10.00}}
\put(633.00,150.00){\circle*{10.00}}
\put(605.00,132.00){$ {\small q_{1}=q_{2}} $}
\put(535.00,132.00){$ {\small q_{3}} $}
\put(535.00,165.00){$ {\small x } $}
\put(625.00,165.00){$ {\small y } $}
\put(480.00,150.00){\line(1,0){210.00}}
\end{picture}

\unitlength=0.6500000pt 
\begin{picture}(300.0,70.00)(150.00,125.00)

\put(250.00,150.00){\circle*{10.00}}
\put(337.00,150.00){\circle*{10.00}}
\put(343.00,150.00){\circle*{10.00}}

\put(245.00,165.00){$ {\small x } $}
\put(335.00,165.00){$ {\small y } $}

\put(315.00,132.00){$ {\small q_{2}=q_{3}} $}
\put(245.00,132.00){$ {\small q_{1}} $}

\put(190.00,150.00){\line(1,0){210.00}}
\put(420.00,147.00){$ = \,\, e^{\imath \pi s} $}

\put(630.00,150.00){\circle*{10.00}}

\put(537.00,150.00){\circle*{10.00}}
\put(543.00,150.00){\circle*{10.00}}

\put(515.00,132.00){$ {\small q_{2}=q_{3}} $}
\put(625.00,132.00){$ {\small q_{1}} $}

\put(535.00,165.00){$ {\small x } $}
\put(625.00,165.00){$ {\small y } $}
\put(480.00,150.00){\line(1,0){210.00}}
\end{picture}

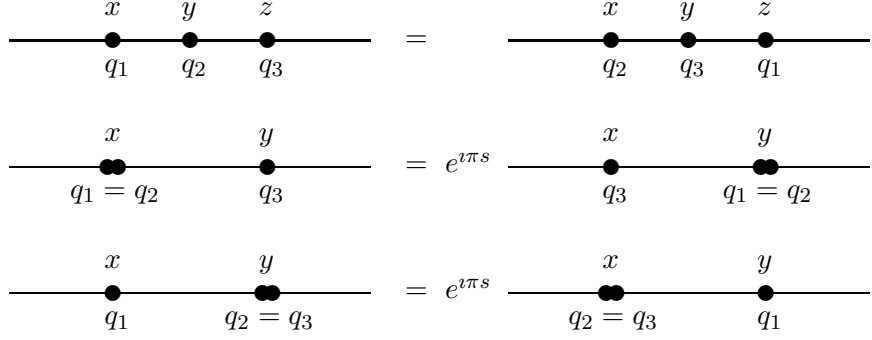
\smallskip \noindent {\small Figure 2: Anyonic exchange factors for the
3-particle scattering in the $A_2$-model.}

Notice that the first case in figure 2, leading to a bosonic exchange
possesses an analogue in the undeformed case. This process can be viewed in
two alternative ways, either corresponding to two consecutive two particle
exchanges, i.e. $1\leftrightarrow 2$ and subsequently $1\leftrightarrow 3$,
or equivalently to a simultaneous three particle scattering process that is
the ordering $123$ goes to $231$ in one scattering event. This is the
typical factorization of an $n$-particle scattering process into a sequence
of two-particle scatterings encountered in integrable models, see e.g. \cite%
{WittenShankar}. In fact, as this feature is so central it is often used
synonymously with integrability. In our deformed model we encounter new
possibilities, namely that a compound particle can exist in the first place
and then also scatter with a single particle; giving rise to anyonic
exchange factors in this case.

\subsubsection{Deformed $A_{3}$-models}

\paragraph{Based on $\mathcal{PT}$-symmetrically deformed Coxeter group
factors}

In this case the potential and $\tilde{z}$ are computed from the inner
products of all $6$ roots in $\tilde{\Delta}_{A_{3}}^{+}$ with $q$. Taking
the simple roots in the standard four dimensional representation $\alpha
_{1}=\{1,-1,0,0\}$, $\alpha _{2}=\{0,1,-1,0\}$, $\alpha _{3}=\{0,0,1,-1\}$,
we evaluate with (\ref{a3}) and (\ref{a3sol})

\begin{eqnarray}
\tilde{\alpha}_{1}\cdot q &=&q_{43}+\cosh \varepsilon (q_{12}+q_{34})-\imath 
\sqrt{2\cosh \varepsilon }\sinh \frac{\varepsilon }{2}(q_{13}+q_{24}),
\label{p1} \\
\tilde{\alpha}_{2}\cdot q &=&q_{23}(2\cosh \varepsilon -1)+\imath 2\sqrt{%
2\cosh \varepsilon }\sinh \frac{\varepsilon }{2}q_{14}, \\
\tilde{\alpha}_{3}\cdot q &=&q_{21}+\cosh \varepsilon (q_{12}+q_{34})-\imath 
\sqrt{2\cosh \varepsilon }\sinh \frac{\varepsilon }{2}(q_{13}+q_{24}), \\
\tilde{\alpha}_{4}\cdot q &=&q_{42}+\cosh \varepsilon (q_{13}+q_{24})+\imath 
\sqrt{2\cosh \varepsilon }\sinh \frac{\varepsilon }{2}(q_{12}+q_{34}), \\
\tilde{\alpha}_{5}\cdot q &=&q_{31}+\cosh \varepsilon (q_{13}+q_{24})+\imath 
\sqrt{2\cosh \varepsilon }\sinh \frac{\varepsilon }{2}(q_{12}+q_{34}), \\
\tilde{\alpha}_{6}\cdot q &=&q_{14}(2\cosh \varepsilon -1)-\imath \sqrt{%
2\cosh \varepsilon }\sinh \frac{\varepsilon }{2}q_{23}.  \label{p6}
\end{eqnarray}
Once again the last term in the potential (\ref{adC}) resulting from these
products is no longer singular in any two particle exchange. However, in
this case it could become singular in two simultaneous two-particle
scattering processes, e.g. $q_{14}=q_{23}=0$. Once again we may realize the $%
\mathcal{PT}$-symmetry constructed for the $\tilde{\alpha}$ 
\begin{eqnarray}
\sigma _{-}^{\varepsilon } &:&\quad \tilde{\alpha}_{1}\rightarrow -\tilde{%
\alpha}_{1}\text{, }\tilde{\alpha}_{2}\rightarrow \tilde{\alpha}_{6}\text{, }%
\tilde{\alpha}_{3}\rightarrow -\tilde{\alpha}_{3}\text{, }\tilde{\alpha}%
_{4}\rightarrow \tilde{\alpha}_{5}\text{, }\tilde{\alpha}_{5}\rightarrow 
\tilde{\alpha}_{4}\text{, }\tilde{\alpha}_{6}\rightarrow \tilde{\alpha}_{2},
\\
\sigma _{+}^{\varepsilon } &:&\quad \tilde{\alpha}_{1}\rightarrow \tilde{%
\alpha}_{4}\text{, }\tilde{\alpha}_{2}\rightarrow -\tilde{\alpha}_{2}\text{, 
}\tilde{\alpha}_{3}\rightarrow \tilde{\alpha}_{5}\text{, }\tilde{\alpha}%
_{4}\rightarrow \tilde{\alpha}_{1}\text{, }\tilde{\alpha}_{5}\rightarrow 
\tilde{\alpha}_{3}\text{, }\tilde{\alpha}_{6}\rightarrow \tilde{\alpha}_{6},
\end{eqnarray}
also in the dual space 
\begin{eqnarray}
\sigma _{-}^{\varepsilon } &:&\quad q_{1}\rightarrow q_{2}\text{, }%
q_{2}\rightarrow q_{1}\text{, }q_{3}\rightarrow q_{4}\text{, }%
q_{4}\rightarrow q_{3}\text{,~}\imath \rightarrow -\imath , \\
\sigma _{+}^{\varepsilon } &:&\quad q_{1}\rightarrow q_{1}\text{, }%
q_{2}\rightarrow q_{3}\text{, }q_{3}\rightarrow q_{2}\text{, }%
q_{4}\rightarrow q_{4}\text{,~}\imath \rightarrow -\imath .
\end{eqnarray}
As in the $A_{2}$-case $\tilde{z}$ will not vanish when two $q$s coincide,
but once again we may pick up nontrivial exchange factors when involving all
particles in the model in the scattering process. We observe 
\begin{eqnarray}
\sigma _{-}^{\varepsilon }\tilde{z}(q_{1},q_{2},q_{3},q_{4}) &=&\tilde{z}%
^{\ast }(q_{2},q_{1},q_{4},q_{3})=\tilde{z}(q_{1},q_{2},q_{3},q_{4}),
\label{a4z1} \\
\sigma _{+}^{\varepsilon }\tilde{z}(q_{1},q_{2},q_{3},q_{4}) &=&\tilde{z}%
^{\ast }(q_{1},q_{3},q_{2},q_{4})=-\tilde{z}(q_{1},q_{2},q_{3},q_{4}).
\label{a4z2}
\end{eqnarray}
Combining (\ref{a4z1}) and (\ref{a4z2}) then yields 
\begin{equation}
\tilde{z}(q_{1},q_{2},q_{3},q_{4})=-\tilde{z}(q_{2},q_{4},q_{1},q_{3}),
\end{equation}
and therefore we will encounter nontrivial exchange factors in a $4$%
-particle scattering process 
\begin{equation}
\psi (q_{1},q_{2},q_{3},q_{4})=e^{\imath \pi s}\psi
(q_{2},q_{4},q_{1},q_{3}).
\end{equation}
We depict various possibilities in figure 3.

\unitlength=0.6500000pt 
\begin{picture}(300.0,70.00)(150.00,125.00)

\put(200.00,150.00){\circle*{10.00}}
\put(250.00,150.00){\circle*{10.00}}
\put(300.00,150.00){\circle*{10.00}}
\put(350.00,150.00){\circle*{10.00}}

\put(195.00,165.00){$ {\small w } $}
\put(245.00,165.00){$ {\small x } $}
\put(295.00,165.00){$ {\small y } $}
\put(345.00,165.00){$ {\small z } $}

\put(195.00,132.00){$ {\small q_{1}} $}
\put(245.00,132.00){$ {\small q_{2}} $}
\put(295.00,132.00){$ {\small q_{3}} $}
\put(345.00,132.00){$ {\small q_{4}} $}

\put(150.00,150.00){\line(1,0){250.00}}
\put(420.00,147.00){$ = \,\, e^{\imath \pi s} $}

\put(530.00,150.00){\circle*{10.00}}
\put(580.00,150.00){\circle*{10.00}}
\put(630.00,150.00){\circle*{10.00}}
\put(680.00,150.00){\circle*{10.00}}

\put(525.00,165.00){$ {\small w } $}
\put(575.00,165.00){$ {\small x } $}
\put(625.00,165.00){$ {\small y } $}
\put(675.00,165.00){$ {\small z } $}

\put(525.00,132.00){$ {\small q_{2}} $}
\put(575.00,132.00){$ {\small q_{4}} $}
\put(625.00,132.00){$ {\small q_{1}} $}
\put(675.00,132.00){$ {\small q_{3}} $}

\put(480.00,150.00){\line(1,0){250.00}}
\end{picture}

\unitlength=0.6500000pt 
\begin{picture}(300.0,70.00)(150.00,125.00)
\put(200.00,150.00){\circle*{10.00}}
\put(272.00,150.00){\circle*{10.00}}
\put(278.00,150.00){\circle*{10.00}}
\put(350.00,150.00){\circle*{10.00}}

\put(195.00,165.00){$ {\small x } $}
\put(270.00,165.00){$ {\small y } $}
\put(345.00,165.00){$ {\small z } $}

\put(250.00,132.00){$ {\small q_{2}=q_{3} } $}
\put(195.00,132.00){$ {\small q_{1}} $}
\put(345.00,132.00){$ {\small q_{4}} $}

\put(150.00,150.00){\line(1,0){250.00}}

\put(420.00,147.00){$ = \,\, e^{\imath \pi s} $}

\put(530.00,150.00){\circle*{10.00}}
\put(602.00,150.00){\circle*{10.00}}
\put(607.00,150.00){\circle*{10.00}}
\put(680.00,150.00){\circle*{10.00}}

\put(525.00,165.00){$ {\small x } $}
\put(600.00,165.00){$ {\small y } $}
\put(675.00,165.00){$ {\small z } $}

\put(523.00,132.00){$ {\small q_{2}  } $}
\put(580.00,132.00){$ {\small q_{1}=q_{4}  } $}
\put(673.00,132.00){$ {\small q_{3}  } $}

\put(480.00,150.00){\line(1,0){250.00}}
\end{picture}

\unitlength=0.6500000pt 
\begin{picture}(300.0,70.00)(150.00,125.00)

\put(197.00,150.00){\circle*{10.00}}
\put(203.00,150.00){\circle*{10.00}}
\put(347.00,150.00){\circle*{10.00}}
\put(353.00,150.00){\circle*{10.00}}

\put(195.00,165.00){$ {\small x } $}
\put(345.00,165.00){$ {\small y } $}

\put(173.00,132.00){$ {\small q_{1}=q_{2}} $}
\put(325.00,132.00){$ {\small q_{3}=q_{4}} $}

\put(150.00,150.00){\line(1,0){250.00}}
\put(420.00,147.00){$ = \,\, e^{\imath \pi s} $}

\put(527.00,150.00){\circle*{10.00}}
\put(533.00,150.00){\circle*{10.00}}
\put(677.00,150.00){\circle*{10.00}}
\put(683.00,150.00){\circle*{10.00}}

\put(525.00,165.00){$ {\small x } $}
\put(675.00,165.00){$ {\small y } $}

\put(503.00,132.00){$ {\small q_{1}=q_{3}  } $}
\put(655.00,132.00){$ {\small q_{2}=q_{4}  } $}

\put(480.00,150.00){\line(1,0){250.00}}
\end{picture}

\unitlength=0.6500000pt 
\begin{picture}(300.0,70.00)(150.00,125.00)
\put(195.00,150.00){\circle*{10.00}}
\put(200.00,150.00){\circle*{10.00}}
\put(205.00,150.00){\circle*{10.00}}
\put(350.00,150.00){\circle*{10.00}}
\put(195.00,165.00){$ {\small x } $}
\put(345.00,165.00){$ {\small y } $}
\put(153.00,132.00){$ {\small q_{1}=q_{2}=q_{3} } $}
\put(345.00,132.00){$ {\small q_{4}} $}
\put(150.00,150.00){\line(1,0){250.00}}
\put(420.00,147.00){$ = $}
\put(530.00,150.00){\circle*{10.00}}
\put(675.00,150.00){\circle*{10.00}}
\put(680.00,150.00){\circle*{10.00}}
\put(685.00,150.00){\circle*{10.00}}
\put(525.00,165.00){$ {\small x } $}
\put(675.00,165.00){$ {\small y } $}
\put(523.00,132.00){$ {\small q_{4}  } $}
\put(635.00,132.00){$ {\small q_{1}=q_{2}=q_{3}} $}
\put(480.00,150.00){\line(1,0){250.00}}
\end{picture}

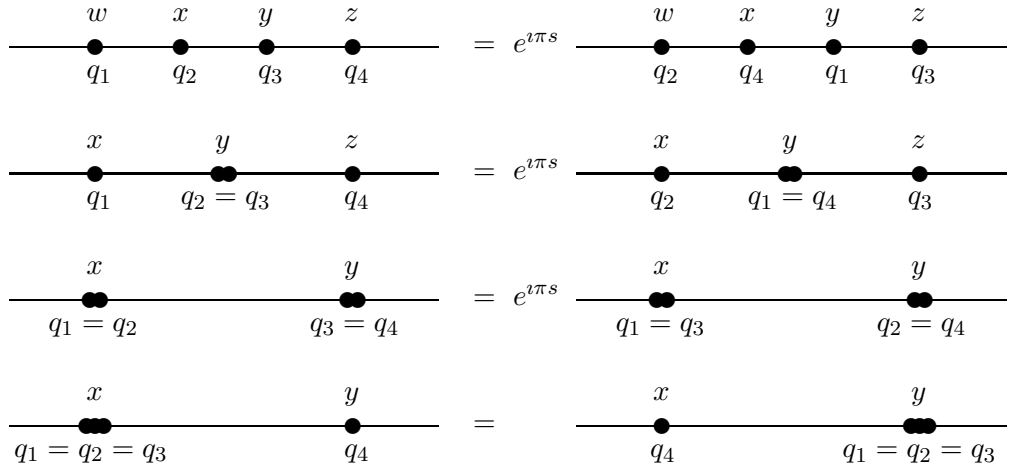
\smallskip \noindent {\small Figure 3: Anyonic exchange factors for the
4-particle scattering in the $A_3$-model.}

As in the previous case we encounter several possibilities which have no
counterpart in the undeformed case.

\paragraph{\noindent Based on $\mathcal{CT}$-symmetrically deformed longest
element}

We keep now the representation for the simple roots, but use the
construction for the deformed roots as provided in the second part of
section \ref{deltaa3}. The potential is obtained again by computing 
\begin{eqnarray}
\tilde{\alpha}_{1}\cdot q &=&\cosh \varepsilon q_{12}+\imath \sinh
\varepsilon q_{34}, \\
\tilde{\alpha}_{2}\cdot q &=&\cosh ^{2}\frac{\varepsilon }{2}q_{23}-\sinh
^{2}\frac{\varepsilon }{2}q_{14}+\frac{\imath }{2}\sinh \varepsilon
(q_{12}+q_{43}), \\
\tilde{\alpha}_{3}\cdot q &=&\cosh \varepsilon q_{34}+\imath \sinh
\varepsilon q_{21}, \\
\tilde{\alpha}_{4}\cdot q &=&\cosh \varepsilon q_{13}-\imath \sinh
\varepsilon q_{24}, \\
\tilde{\alpha}_{5}\cdot q &=&\cosh \varepsilon q_{24}+\imath \sinh
\varepsilon q_{13}, \\
\tilde{\alpha}_{6}\cdot q &=&\cosh ^{2}\frac{\varepsilon }{2}q_{14}+\sinh
^{2}\frac{\varepsilon }{2}q_{23}+\frac{\imath }{2}\sinh \varepsilon
(q_{21}+q_{34}).
\end{eqnarray}
Clearly the potential is different from the one resulting from (\ref{p1})-(%
\ref{p6}). Despite the fact that it is a simpler potential, it can not be
solved analogously to the previous case since the crucial relations (\ref{11}%
)-(\ref{14}) no longer hold.

\subsubsection{The deformed $F_{4}$-model}

In order to unravel any features which might differ in the non-simply laced
case, which is usually the case, we also present her one example for such a
model. To allow a direct comparison with the previous 4-particle case, we
have selected $F_{4}$. The positive root space $\tilde{\Delta}_{F_{4}}^{+}$
contains now $24$ root. Taking the simple roots in the standard four
dimensional representation $\alpha _{1}=\{0,1,-1,0\}$, $\alpha
_{2}=\{0,0,1,-1\}$, $\alpha _{3}=\{0,0,0,1\}$ and\ $\alpha
_{4}=\{1/2,-1/2,-1/2,-1/2\}$ we compute the following factorization for $%
\tilde{z}$, with each factor corresponding to one of the 24 products $\tilde{%
\alpha}_{i}\cdot q$

\begin{eqnarray*}
&&\left( q_{1}\cosh \varepsilon +\imath \sinh \varepsilon q_{4}\right)
\left( q_{2}\cosh \varepsilon -\imath \sinh \varepsilon q_{3}\right) \left(
q_{3}\cosh \varepsilon +\imath \sinh \varepsilon q_{2}\right) \left(
q_{4}\cosh \varepsilon -\imath \sinh \varepsilon q_{1}\right) \\
&&\times \left( q_{12}\cosh \varepsilon +\imath \sinh \varepsilon \hat{q}%
_{34}\right) \left( q_{14}\cosh \varepsilon +\imath \sinh \varepsilon \hat{q}%
_{14}\right) \left( q_{34}\cosh \varepsilon +\imath \sinh \varepsilon \hat{q}%
_{12}\right) \\
&&\times \left( q_{23}\cosh \varepsilon -\imath \sinh \varepsilon \hat{q}%
_{23}\right) \left( \hat{q}_{13}\cosh \varepsilon +\imath \sinh \varepsilon 
\hat{q}_{24}\right) \left( \hat{q}_{24}\cosh \varepsilon -\imath \sinh
\varepsilon \hat{q}_{13}\right) \\
&&\times \left( \hat{q}_{34}\cosh \varepsilon -\imath \sinh \varepsilon
q_{12}\right) \left( \hat{q}_{23}\cosh \varepsilon +\imath \sinh \varepsilon 
\hat{q}_{23}\right) \left( \hat{q}_{12}\cosh \varepsilon -\imath \sinh
\varepsilon \hat{q}_{34}\right) \\
&&\times \left( \hat{q}_{14}\cosh \varepsilon -\imath \sinh \varepsilon
q_{14}\right) \left( q_{24}\cosh \varepsilon +\imath \sinh \varepsilon
q_{13}\right) \left( q_{13}\cosh \varepsilon -\imath \sinh \varepsilon
q_{24}\right) \\
&&\times \left[ \frac{\hat{q}_{12}+\hat{q}_{34}}{2}\cosh \varepsilon -\frac{%
\imath }{2}\sinh \varepsilon (q_{12}+q_{34})\right] \left[ \frac{\hat{q}%
_{12}-q_{34}}{2}\cosh \varepsilon -\frac{\imath }{2}\sinh \varepsilon (\hat{q%
}_{12}+q_{34})\right] \\
&&\times \left[ \frac{q_{12}-\hat{q}_{34}}{2}\cosh \varepsilon +\frac{\imath 
}{2}\sinh \varepsilon (q_{12}+\hat{q}_{34})\right] \left[ \frac{\hat{q}_{12}-%
\hat{q}_{34}}{2}\cosh \varepsilon +\frac{\imath }{2}\sinh \varepsilon
(q_{12}-q_{34})\right] \\
&&\times \left[ \frac{q_{12}+\hat{q}_{34}}{2}\cosh \varepsilon -\frac{\imath 
}{2}\sinh \varepsilon (q_{12}-\hat{q}_{34})\right] \left[ \frac{q_{12}-q_{34}%
}{2}\cosh \varepsilon -\frac{\imath }{2}\sinh \varepsilon (\hat{q}_{12}-\hat{%
q}_{34})\right] \\
&&\times \left[ \frac{q_{12}+q_{34}}{2}\cosh \varepsilon +\frac{\imath }{2}%
\sinh \varepsilon (\hat{q}_{12}+\hat{q}_{34})\right] \left[ \frac{%
q_{12}+q_{34}}{2}\cosh \varepsilon +\frac{\imath }{2}\sinh \varepsilon (\hat{%
q}_{12}-q_{34})\right] ,
\end{eqnarray*}
where we used the abbreviation $\hat{q}_{ij}:=q_{i}+q_{j}$. Once again,
several singularities have disappeared through the deformation. The $%
\mathcal{PT}$-symmetry constructed for the simple deformed roots $\tilde{%
\alpha}$ 
\begin{eqnarray}
\sigma _{-}^{\varepsilon } &:&\quad \tilde{\alpha}_{1}\rightarrow -\tilde{%
\alpha}_{1}\text{, }\tilde{\alpha}_{2}\rightarrow \tilde{\alpha}_{1}+\tilde{%
\alpha}_{2}+2\tilde{\alpha}_{3}\text{, }\tilde{\alpha}_{3}\rightarrow -%
\tilde{\alpha}_{3}\text{, }\tilde{\alpha}_{4}\rightarrow \tilde{\alpha}_{3}+%
\tilde{\alpha}_{4}, \\
\sigma _{+}^{\varepsilon } &:&\quad \tilde{\alpha}_{1}\rightarrow \tilde{%
\alpha}_{1}+\tilde{\alpha}_{2}\text{, }\tilde{\alpha}_{2}\rightarrow -\tilde{%
\alpha}_{2}\text{, }\tilde{\alpha}_{3}\rightarrow \tilde{\alpha}_{2}+\tilde{%
\alpha}_{3}+\tilde{\alpha}_{4}\text{, }\tilde{\alpha}_{4}\rightarrow -\tilde{%
\alpha}_{4},
\end{eqnarray}
is now realized in the dual space as 
\begin{eqnarray}
\sigma _{-}^{\varepsilon } &:&\quad q_{1}\rightarrow q_{1}\text{, }%
q_{2}\rightarrow q_{3}\text{, }q_{3}\rightarrow q_{2}\text{, }%
q_{4}\rightarrow -q_{4}\text{,~}\imath \rightarrow -\imath , \\
\sigma _{+}^{\varepsilon } &:&\quad q_{1}\rightarrow \frac{1}{2}%
(q_{1}+q_{2}+q_{3}+q_{4})\text{, }q_{2}\rightarrow \frac{1}{2}%
(q_{1}+q_{2}-q_{3}-q_{4})\text{, } \\
\qquad &&\quad q_{3}\rightarrow \frac{1}{2}(q_{1}-q_{2}-q_{3}+q_{4})\text{, }%
q_{4}\rightarrow \frac{1}{2}(q_{1}-q_{2}+q_{3}-q_{4})\text{,~}\imath
\rightarrow -\imath .~
\end{eqnarray}
Now we observe 
\begin{eqnarray}
\sigma _{-}^{\varepsilon }\tilde{z}(q_{1},q_{2},q_{3},q_{4}) &=&\tilde{z}%
^{\ast }(q_{1},q_{3},q_{2},-q_{4})=\tilde{z}(q_{1},q_{2},q_{3},q_{4}),
\label{f4z1} \\
\sigma _{+}^{\varepsilon }\tilde{z}(q_{1},q_{2},q_{3},q_{4}) &=&\tilde{z}%
^{\ast }\left[ \frac{\hat{q}_{12}+\hat{q}_{34}}{2},\frac{\hat{q}_{12}-\hat{q}%
_{34}}{2},\frac{q_{12}-q_{34}}{2},\frac{q_{12}+q_{34}}{2}\right] =\tilde{z}%
(q_{1},q_{2},q_{3},q_{4}).~~~~~~  \label{f4z2}
\end{eqnarray}
A consequence of this we find the symmetry 
\begin{equation}
\psi (q_{1},q_{2},q_{3},q_{4})=\psi (\frac{\hat{q}_{13}+q_{24}}{2},\frac{%
\hat{q}_{13}-q_{34}}{2},\frac{q_{13}-\hat{q}_{24}}{2},\frac{q_{13}+\hat{q}%
_{24}}{2}),
\end{equation}
which gives rise to new possibilities neither encountered in the undeformed
case nor in the deformed $A_{3}$-case.

\section{Conclusions}

As a particular element in the Coxeter group we have selected the
involutions $\sigma _{\pm }$, $\omega _{0}\in \mathcal{W}$ and deformed them
in an antilinear manner. For each construction we have set up sets of
constraining equations (\ref{const}) and (\ref{constw}) for the deformation
matrix $\theta _{\varepsilon }$ and solved them case-by-case. This matrix
then yields via (\ref{roots}) the deformed simple roots and with the help of
the deformed Coxeter element $\sigma ^{\varepsilon }$ we have constructed
the remaining nonsimple roots and thus the entire deformed root space $%
\tilde{\Delta}(\varepsilon )$. Depending on the type of construction this
space remains invariant under the action of the new elements $\sigma _{\pm
}^{\varepsilon }$, $\omega _{0}^{\varepsilon }$, i.e. an antilinear
transformation, and in the former case also under the action of $\sigma
^{\varepsilon }$. The construction based on the deformation of $\omega _{0}$
is more restrictive from the very onset as it can only be applied to $%
A_{\ell }$, $D_{2\ell }$ and $E_{6}$. In addition, the resulting deformed
root space is only invariant under the action of $\sigma ^{\varepsilon }$
when the construction coincides with the one based on the deformation of the 
$\sigma _{\pm }$. The latter construction is more general as it can be
applied to all Coxeter groups, albeit it does not always lead to nontrivial
solutions. We have demonstrated that for the cases for which one can only
find real solutions one may still find complex solutions of different type
by means of the folding procedure.

Clearly there are various open issues with regard to the mathematical
framework. We have for instance not constructed in all cases the most
general deformation possible, even for the constraints we have provided.
Besides the closed generic formulae we found for some infinite series it
might be possible to construct them also for the missing ones. Furthermore,
it would be interesting to relax some of the constraints we impose on the
deformation matrix and construct new types of solutions, especially in those
cases for which we showed that no complex solutions exist. Finally it would
also be interesting to investigate the possibility to deform different types
of involutions besides the $\sigma _{\pm }$ and $\omega _{0}$ presented here.

We argued that the deformed root systems may be used to define new types of
physical models, whose formulation is based on root systems, such as Toda
lattice theories, affine or conformal Toda field theories,
Calogero-Moser-Sutherland models etc. By construction the Hamiltonians
related to all these models will be invariant under some antilinear
transformation, $\sigma _{\pm }^{\varepsilon }$ or $\omega _{0}^{\varepsilon
}$, and therefore have a strong likelihood to be physically meaningful with
real spectra. We have exploited this possibility in some detail for Calogero
models. We have constructed specific solutions to the Schr\"{o}dinger
equation leading to the same real energy spectrum as in the undeformed case,
but to different wavefunctions. A particular interesting new feature of the
new models is that they give rise to anyonic exchange factors of various
types of particle exchanges which have no analogue in the undeformed case.
Our solution procedure relies entirely on the validity of the identities
presented in appendix A, for which we have presented strong evidence on
various case-by-case studies. For completeness it would be desirable to have
some generic proofs for them. Most interesting from a physical point of view
would be to find further solutions to the Schr\"{o}dinger equation besides
the $l=0$ ones. As in the $G_{2}$ and $A_{2}$ case this will most likely
give rise to a different energy spectrum similarly to the undeformed case.
Related to this issue is the interesting question of how the Hermitian
counterpart or possibly counterparts to $\mathcal{H}_{adC}(p,q)$ would look
like. Naturally one would also like to answer the question of whether the
deformed models are still integrable, which is especially interesting in the
light of the comments made at the end of section 3.2.1.

\medskip

\noindent \textbf{Acknowledgments:} M.S. is supported by EPSRC. A.F. would
like to thank the National Institute for Theoretical Physics of South Africa
and the Stellenbosch Institute for Advanced Studies for their kind
hospitality and financial support, where parts of this work were carried
out. Special thanks go to Frederik Scholtz and Laure Gouba for stimulating
discussions.

\newpage

\appendix

\section{Identities}

We assemble here the crucial identities for the derivation of the radial
part of the Schr\"{o}dinger equation (\ref{radial}). Underlying are the
generic relations which only involve roots and the dynamical variables $%
q=\{q_{1},\ldots ,q_{n}\}$ 
\begin{eqnarray}
\sum\limits_{\alpha ,\beta \in \Delta ^{+}}\frac{\alpha \cdot \beta }{%
(\alpha \cdot q)(\beta \cdot q)} &=&\sum\limits_{\alpha \in \Delta ^{+}}%
\frac{\alpha ^{2}}{(\alpha \cdot q)^{2}},  \label{11} \\
\sum\limits_{\alpha ,\beta \in \Delta ^{+}}(\alpha \cdot \beta )\frac{%
(\alpha \cdot q)}{(\beta \cdot q)} &=&\frac{\hat{h}h\ell }{2}t_{\ell },
\label{12} \\
\sum\limits_{\alpha ,\beta \in \Delta ^{+}}\left( \alpha \cdot \beta \right)
(\alpha \cdot q)(\beta \cdot q) &=&\hat{h}t_{\ell }\sum\limits_{\alpha \in
\Delta ^{+}}(\alpha \cdot q)^{2},  \label{13} \\
\sum\limits_{\alpha \in \Delta ^{+}}\alpha ^{2} &=&\ell \hat{h}t_{\ell }.
\label{14}
\end{eqnarray}%
At present we do not have a generic proof for these relations. A large
amount of evidence on a case-by-case basis for the first identity was
already provided in \cite{AF}, albeit no case independent generic proof.
Here we have verified (\ref{12}) and (\ref{13}) for a substantial number of
Coxeter groups. Denoting by $n_{s}$, $\alpha _{s}^{2}$ and $n_{l}$, $\alpha
_{l}^{2}$ the number and length of the short and long roots, respectively, (%
\ref{14}) follows from 
\begin{equation}
\sum\limits_{\alpha \in \Delta ^{+}}\alpha ^{2}=\frac{n_{s}}{2}\alpha
_{s}^{2}+\frac{n_{l}}{2}\alpha _{l}^{2}=\frac{\alpha _{l}^{2}}{2}\left( n_{s}%
\frac{\alpha _{s}^{2}}{\alpha _{l}^{2}}+n_{l}\right) =\ell \hat{h}t_{\ell },
\end{equation}%
where we used $n_{s}\alpha _{s}^{2}/\alpha _{l}^{2}+n_{l}=\ell \hat{h}$,
which can be found for instance in \cite{Goddard:1986bp} and $\alpha
_{l}^{2}=2t_{\ell }$.

Accepting the relations (\ref{11})-(\ref{14}) the identities involving
derivatives of $r$ and $z$ are easily derived. From (\ref{rz}) follows

\begin{equation}
\frac{\partial z}{\partial q_{i}}=z\sum\limits_{\alpha \in \Delta ^{+}}\frac{%
\alpha ^{i}}{(\alpha \cdot q)}\qquad \text{and\qquad }\frac{\partial r}{%
\partial q_{i}}=\frac{1}{r\hat{h}t_{\ell }}\sum\limits_{\alpha \in \Delta
^{+}}(\alpha \cdot q)\alpha ^{i}\text{.}  \label{drz}
\end{equation}
Multiplying them and summing over the dynamical variables gives 
\begin{eqnarray}
\sum\limits_{i=1}^{n}\left( \frac{\partial z}{\partial q_{i}}\right) ^{2}
&=&z^{2}\sum\limits_{\alpha ,\beta \in \Delta ^{+}}\frac{\alpha \cdot \beta 
}{(\alpha \cdot q)(\beta \cdot q)}=z^{2}\sum\limits_{\alpha \in \Delta ^{+}}%
\frac{\alpha ^{2}}{(\alpha \cdot q)^{2}},  \label{21} \\
\sum\limits_{i=1}^{n}\frac{\partial z}{\partial q_{i}}\frac{\partial r}{%
\partial q_{i}} &=&\frac{z}{\hat{h}t_{\ell }r}\sum\limits_{\alpha ,\beta \in
\Delta ^{+}}(\alpha \cdot \beta )\frac{(\alpha \cdot q)}{(\beta \cdot q)}=%
\frac{h\ell }{2}\frac{z}{r},  \label{22} \\
\sum\limits_{i=1}^{n}\left( \frac{\partial r}{\partial q_{i}}\right) ^{2} &=&%
\frac{1}{r^{2}\hat{h}^{2}t_{\ell }^{2}}\sum\limits_{\alpha ,\beta \in \Delta
^{+}}\left( \alpha \cdot \beta \right) (\alpha \cdot q)(\beta \cdot q)=1,
\label{23}
\end{eqnarray}
where we have used (\ref{11}) in (\ref{21}), (\ref{12}) in (\ref{22}) and (%
\ref{13}) in (\ref{23}). Furthermore we need the sums over the second order
derivatives. From (\ref{drz}) we obtain with the help of (\ref{11}) and (\ref%
{12}) the two relations 
\begin{eqnarray}
\sum\limits_{i=1}^{n}\frac{\partial ^{2}z}{\partial q_{i}^{2}} &=&z\left(
\sum\limits_{\alpha ,\beta \in \Delta ^{+}}\frac{\alpha \cdot \beta }{%
(\alpha \cdot q)(\beta \cdot q)}-\sum\limits_{\alpha \in \Delta ^{+}}\frac{%
\alpha ^{2}}{(\alpha \cdot q)^{2}}\right) =0,  \label{wichtig} \\
\sum\limits_{i=1}^{n}\frac{\partial ^{2}r}{\partial q_{i}^{2}} &=&\frac{1}{r%
\hat{h}t_{\ell }}\sum\limits_{\alpha \in \Delta ^{+}}\alpha ^{2}-\frac{1}{%
r^{3}\hat{h}t_{\ell }}\sum\limits_{\alpha ,\beta \in \Delta ^{+}}\left(
\alpha \cdot \beta \right) (\alpha \cdot q)(\beta \cdot q)=\frac{\ell -1}{r}.
\label{auch}
\end{eqnarray}

\section{Case-by-case data}

For convenience we present in this appendix some numerical data for
individual Coxeter groups. We present the values for the Coxeter number $h$
defined as the total number of roots divided by the rank, the order of the
Coxeter element $\sigma $ or $1+\sum\nolimits_{i=1}^{\ell }n_{i}$ when the
highest root is expressed in terms of simple roots as $\sum\nolimits_{i=1}^{%
\ell }n_{i}\alpha _{i}$. The dual Coxeter number is defined in the same way
as the Coxeter number for the situation in which the arrows on the affine
Diagram have been reversed. The exponents $s_{n}$ are related to the
eigenvalues of the Coxeter element as defined in (\ref{exp}) and $t_{\ell }$
is the $\ell $-th symmetrizer of the incidence matrix $I$ defined by means
of the relation $I_{ij}t_{j}=t_{i}I_{ij}$.

\begin{center}
\begin{tabular}{||l||c|c|l|l||}
\hline\hline
$\mathcal{W}$ & $h$ & $\hat{h}$ & $s_{n}$ & $t_{\ell }$ \\ \hline\hline
$A_{\ell }$ & $\ell +1$ & $\ell +1$ & $1,2,3,...,\ell $ & $1$ \\ \hline
$B_{\ell }$ & $2\ell $ & $2\ell -1$ & $1,3,5,...,2\ell -1$ & $1$ \\ \hline
$C_{\ell }$ & $2\ell $ & $\ell +1$ & $1,3,5,...,2\ell -1$ & $2$ \\ \hline
$D_{\ell }$ & $2\ell -2$ & $2\ell -2$ & $1,3,...,\ell -1,\ldots ,2\ell -3$ & 
$1$ \\ \hline
$E_{6}$ & $12$ & $12$ & $1,4,5,7,8,11$ & $1$ \\ \hline
$E_{7}$ & $18$ & $18$ & $1,5,7,9,11,13,17$ & $1$ \\ \hline
$E_{8}$ & $30$ & $30$ & $1,7,11,13,17,19,23,29$ & $1$ \\ \hline
$F_{4}$ & $12$ & $9$ & $1,5,7,11$ & $1$ \\ \hline
$G_{2}$ & $6$ & $4$ & $1,5$ & $3$ \\ \hline
$H_{3}$ & $10$ & $10$ & $1,5,9$ & $1$ \\ \hline
\end{tabular}
\end{center}

\noindent Table 1: Coxeter number $h$, dual Coxeter number $\hat{h}$,
exponents $s_{n}$ and $\ell $th symmetrizer $t_{\ell }$.


\end{document}